\documentclass[10pt,final,journal,compsoc]{IEEEtran}
\usepackage[utf8]{inputenc}
\usepackage{graphicx}
\usepackage{amsmath}
\usepackage[hidelinks]{hyperref}
\usepackage{tabularx}
\usepackage{booktabs}
\usepackage{rotating}
\usepackage{multirow}
\usepackage{multicol}
\usepackage[nobreak,space,compress]{cite}

\usepackage{wrapfig}
\usepackage{ragged2e}
\usepackage{xcolor}
\usepackage{comment}
\usepackage{soul}
\usepackage{colortbl}
\usepackage{multirow}
\newcommand{\highlightrow}[1]{\textcolor{black}{\cellcolor{lightgray}#1}}
\usepackage{tikz}
\usepackage{verbatim}
\usetikzlibrary{positioning}
\usepackage{graphicx}
\usepackage{xcolor}
\usepackage{fontawesome5}
\usepackage{makecell}
\usepackage{footnote}
\usepackage{hyperref}
\usepackage{float}
\usepackage{subcaption}
\usepackage{xcolor} % Required for text color
\usepackage{tcolorbox}
\usepackage{caption}
\usepackage{float}
\usepackage{marginnote}

\usepackage[backgroundcolor=white,bordercolor=blue,linecolor=blue]{todonotes}
\newcommand{\ourapproach}{FlakyFix}

\begin{document}

%=================== Author Info Here =================%

\title{FlakyFix: Using Large Language Models for Predicting Flaky Test Fix Categories and Test Code Repair}

\author{
    Sakina Fatima, 
    Hadi Hemmati, and
    Lionel Briand, \IEEEmembership{Fellow, IEEE}
\IEEEcompsocitemizethanks{
    \IEEEcompsocthanksitem Sakina Fatima is with the School of EECS, University of Ottawa, Canada.~E-mail: sfati077@uottawa.ca

    \IEEEcompsocthanksitem Hadi Hemmati is with the Electrical Engineering and Computer Science Department, York University, Canada.~E-mail: hemmati@yorku.ca

    \IEEEcompsocthanksitem Lionel Briand holds shared appointments with the school of EECS, University of Ottawa,  Canada, and the Lero SFI centre for Software Research, University of Limerick, Ireland.~E-mail: lbriand@uottawa.ca
}% <-this % stops an unwanted space
\thanks{}}
\IEEEtitleabstractindextext{%
    \begin{abstract}\justifying
  Flaky tests are problematic because they non-deterministically pass or fail for the same software version under test, causing confusion and wasting development effort. While machine learning models have been used to predict flakiness and its root causes, there is much less work on providing support to fix the problem. To address this gap, in this paper, we focus on predicting the type of fix that is required to remove flakiness and then repair the test code on that basis. We do this for a subset of flaky tests where the root cause of flakiness is in the test itself and not in the production code.
  One key idea is to guide the repair process with additional knowledge about the test's flakiness in the form of its predicted fix category. Thus, we first propose a framework that automatically generates labeled datasets for 13 fix categories and trains models to predict the fix category of a flaky test by analyzing the test code only. Our experimental results using code models and few-shot learning show that we can correctly predict most of the fix categories. To show the usefulness of such fix category labels for automatically repairing flakiness, we augment the prompts of GPT 3.5 Turbo, a Large Language Model (LLM), with such extra knowledge to request repair suggestions. The results show that our suggested fix category labels, complemented with in-context learning, significantly enhance the capability of GPT 3.5 Turbo in generating fixes for flaky tests.~Based on the execution and analysis of a sample of GPT-repaired flaky 
tests, we estimate that a large percentage of such repairs, (roughly between 51\% and 83\%) can be expected to pass. For the failing repaired tests, on average, 16\% of the test code needs to be further changed for them to pass.
    \end{abstract}

}
\maketitle
\IEEEdisplaynontitleabstractindextext
\IEEEpeerreviewmaketitle

\section{Introduction}
% Intro to flaky tests and intro to different ways of fixing the Flaky Tests

% Why flaky tests are big issues

Flaky tests are test cases that non-deterministically fail or pass in different runs, without any changes in the source code being tested. This inconsistency is frustrating and troublesome for developers and Quality Assurance (QA) teams, especially during regression testing. Oftentimes, flakiness results from multiple factors, including variations in configurations used to execute identical code, timing issues influenced by prior executions, race conditions in concurrent systems, and absent external dependencies. 
% flakiness is caused by various factors such as different configurations to run the same code, timing issues with lasting effects of what has been executed before the flaky test, race conditions in concurrent systems, and missing external dependencies.

Flaky tests are a widespread problem in software development and a major challenge to address. With millions of tests in repositories, even a small percentage of flaky tests require significant manual effort to detect and fix. For example, a report from Google shows that they had 1.6M test failures on average each day, and 73K of them (4.56\%) were caused by flaky tests~\cite{luo2014empirical,fallahzadeh2022impact}. Several other companies like Microsoft~\cite{harry2019we}, Facebook~\cite{cordy2019flakime}, Netflix, Mozilla, and Salesforce~\cite{lam2020large} have reported similar experiences, where flaky tests have significantly delayed their deployment process. 

% what is the SOTA now with respect to FT
To detect test flakiness, according to a recent study~\cite{habchi2022qualitative}, practitioners are still relying mainly on reruns and manual inspection. In the literature, state-of-the-art (SOTA) work on handling test flakiness relies on both static techniques~\cite{ahmad2021multi,verdecchia2021know,pontillo2021toward,bell2018deflaker}~and advanced machine learning (ML) models ~\cite{qin2022peeler,fatima2022flakify,gruber2023practical,alshammari2021flakeflagger, pinto2020vocabulary, haben2021replication, camara2021vocabulary,ahmad2020evaluation,parry2021survey} for the automated prediction of flakiness and its causes. However, there is limited work~\cite{parry2022developer} on helping developers and testers fix the flakiness of their test cases. 

% what we are doing
Flakiness in tests can stem from highly diverse reasons~\cite{hashemi2022empirical} and can be resolved through various strategies, as evident from the International Dataset of Flaky Tests (IDoFT)~\cite{InternationalDatasetofFlakyTests}, the largest open-source repository containing Java and Python flaky tests and their respective fixes. Remedies range from altering the test execution order to addressing operating system or implementation dependencies by code modifications in the test class. Our study focuses on repairing flaky tests in a black-box manner (no access to production code) by updating flaky statements within the test case code, which constitutes 10\% (562 tests from accepted pull requests) of the total 5,500 tests in the Java dataset~\cite{InternationalDatasetofFlakyTests}.~While such flaky tests represent a minority in this dataset, techniques to fix them are warranted, thereby serving the important needs of the testing community in charge of writing and evolving the test code.
In addition, studies such as that of Akli et al.~\cite{akli2023flakycat} report that, in their dataset, 70\% of flakiness issues originate in the test code~\cite{luo2014empirical,lam2020study}, further supporting our focus. The distinction between fixing the production code and the test code is crucial. Considering the production code would dramatically increase the costs and complexity of repair, as it would warrant code coverage data collection during test execution for fault localization purposes. This is particularly challenging for complex production code with numerous dependencies and often multiple programming languages~\cite{fatima2022flakify}.
%Additionally, previous industry feedback has highlighted the infeasible expense of achieving comprehensive program coverage~\cite{fatima2022flakify}, a requirement for fixing flakiness in production code.
%In the future, we nevertheless plan to expand our efforts to include white-box flaky test fixes.}

% However, Our study specifically focuses on helping software testers by repairing flaky tests found within the test code, which accounts for 10\% (562 tests from accepted pull requests) out of the total 5500 tests present in the current Java~\cite{InternationalDatasetofFlakyTests} dataset. Generally, The number of flaky tests in a dataset where the flakiness lies in the test code even is a small minority but in the scope of this work we want to address these tests as of now and atleast we want to make sure this small number is correctly fixed. In the future work, we will expand this work from test case to the test class and code under test.  

% our study's scope centers on addressing flaky tests solely within test code, accounting for 10\% (562 tests from accepted pull requests) of the 5500 tests in the current Java dataset. 

More precisely, our goal is to (a) define different categories of fixes that are applied to the test code to remove flakiness; (b) given a flaky test case (and without the production code), predict the most suitable fix category to apply; and (c) incorporate this predicted fix category into an automated flaky test repair approach. 

%The intended purpose of the fix category is to offer guidance to testers, directing them toward potential issues responsible for the flakiness.

%In this paper, we first develop a framework utilizing pre-trained language models to predict the fix category of a flaky test case solely based on the test case code. Our primary aim is to establish this prediction process as a black-box approach, solely reliant on the flaky test code without accessing the production code.

% how this is different 
Our study is different than existing work~\cite{akli2023flakycat} whose focus is on predicting the category of flaky tests with respect to their root causes. Since knowing only root causes, especially at a high-level, does not necessarily help in fixing the test code, our work focuses on predicting fix categories in terms of what part of the test code needs fixing. Therefore, such fix categories are expected to guide the tester towards concrete modifications and thus decrease their manual investigation effort.  
Note that these categories can be useful for both human developers and testers to manually fix flakiness and for tools to automatically and fully repair flaky test cases. In this paper, however, we only report on the latter.

% why and how it is important 
%We chose not to target a fully automated repair mechanism for flaky tests, given that even having a remote chance to achieve high accuracy would require massive training data that are simply not available. Further, in initial studies, using SOTA program repair techniques did not show acceptable results for fixing flakiness, which led us to investigate the prediction of fix categories.  But we believe a semi-automated approach based on such precise guidance is still very beneficial, especially when the number of flaky tests is large and any time saved during manual inspection of the test code will add up to significant savings during the regression testing of large systems.

% how did we asses and what are the results
To achieve the above three goals, our approach consists of three steps: (a) Qualitatively analysing flaky test datasets and proposing a list of fix categories to remove test flakiness, when its root cause is in the test code only;
%Suggesting a classification system to analyze test repairs for automated categorization of flaky tests based on the type of fix they require. 
% proposing a classification system analyzing test fixes for automated fix category labeling,
% proposing a classification of flaky test fixes that is amenable to automated labeling, 
(b) Conducting empirical analyses to explore the use of pre-trained code models in predicting the necessary fix category for addressing identified flaky tests; and 
(c) Evaluating the improvements such categories can provide to an LLM such as GPT when asked to generate fixes for flaky tests.

In step (b) we assess two smaller code models, CodeBERT~\cite{feng2020codebert}, and UniXcoder~\cite{guo2022unixcoder}, which are specifically tailored for smaller datasets and are pre-trained with a reduced parameter set (125 million). We fine-tuned these models with and without Few-Shot Learning (FSL)~\cite{wang2020generalizing}. Considering FSL stems from commonly dealing with small datasets of flaky tests within each fix category, which are typically available for the fine-tuning process.

\begin{figure*}[hbt!]
    \centering
    % \hspace*{-0.5cm}
    \includegraphics[width=1.0\linewidth]{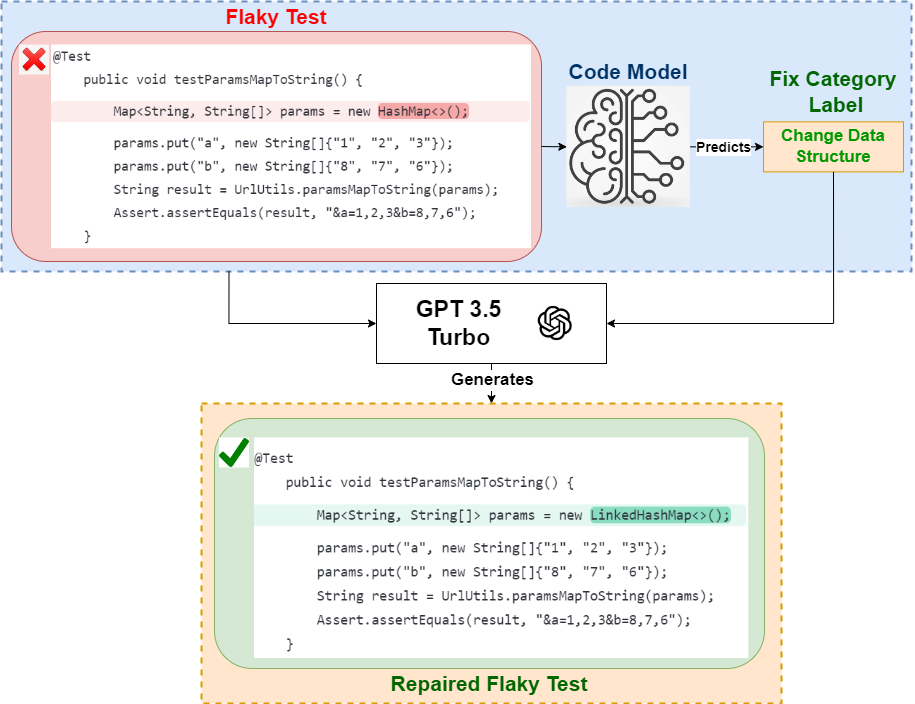}
    \caption{Predicting fix category label and generating repaired flaky tests using large language models.}
    \label{fig:FlakyFixSystem}
    \vspace{-3pt}
\end{figure*}

In step (c), we evaluate the performance of recent Large Language Models (LLMs), namely GPT~\cite{openai2023gpt}, which possess a vast number of model parameters (up to 175 billion in GPT 3.5 Turbo), in analyzing their capability to generate fixes for flaky tests. The steps (b) and (c) are depicted in Figure~\ref{fig:FlakyFixSystem}.
Our findings indicate several key points: (1) Smaller pre-trained code models demonstrate a high accuracy in predicting most fix categories, achieving precision rates ranging from 83\% to 98\% and recall rates from 67\% to 97\%, solely by analyzing the test code; (2) UniXcoder surpasses CodeBERT in predicting fix categories; (3) The inclusion of FSL does not yield significant improvements in the results; (4) Our predicted fix category labels play a significant role in guiding GPT to generate better full fixes for flaky tests. We evaluated this using CodeBLEU~\cite{dong2023codescore} and achieved a mean score of 81\% when using predicted fix category labels in the input prompt, compared to 75\% without them. These results are further enhanced with in-context learning~\cite{liu2021makes}, which we tested on three different fix categories;
(5) Based on the execution and analysis of a sample of GPT-repaired tests, we estimate that a large percentage can be expected to pass, roughly between 46\% and 80\% of them. For the generated tests that fail, the required changes to correct them are limited, with an average of 16\% of the tokens needing modification in the test code.
% list of contributions 
In summary, the contributions of this paper are as follows:
  \begin{itemize}
      \item Definition of a categorization for flaky test fixes. The goal is to provide practical guidance for fixing tests. 
      \item Development of a set of heuristics (search rules) and accompanying open-source scripts to automatically label flaky tests based on their fixes.
      \item Creation of a public dataset of labeled flaky tests for training, using our automated script.  
      \item Definition of a black-box framework for predicting flaky test fix categories using code models (language models that are pre-trained on code) and few-shot learning.
      \item Quantitative and qualitative analysis of two SOTA small code models, with and without FSL, to compare alternative prediction models. 
      \item Automatic generation of fixes for flaky tests using a LLM (GPT 3.5 Turbo), with and without the predicted fix categories, to analyze how these labels enhance the generated fixes. 
      \item Based on the execution of a sample of GPT-generated tests, we rely on CodeBLEU metrics and statistical methods, including bootstrapping and logistic regression, to estimate the pass rate for all generated repaired tests, including those that could not be executed. We also assess the magnitude of the modifications needed to manually fix tests that failed.
  \end{itemize}
  
% organization 
The rest of this paper is organized as follows: Section~\ref{Approach} presents our approach for developing an automated tool for labeling flaky tests with fix categories and predicting such categories using code models, with and without FSL. It also describes how we incorporate such labels as part of prompts for GPT to create a fully automated flaky test repair mechanism. 
Section~\ref{Validation} evaluates our approach, reports experimental results, and discusses their implications.
Section~\ref{Discussion} reports on a qualitative analysis of the two small code models and the LLM to demonstrate and explain their ability to learn from test code and subsequently make predictions and generate fixes. 
Section~\ref{Threats to Validity} analyzes threats to the validity of this study. 
Section~\ref{Related Work} reviews and contrasts related work.
Finally, Section~\ref{Conclusion} concludes the paper and suggests future work.

\section{Approach}\label{Approach}
% Black  Box Flaky Test Fixes Categorization and Prediction
The approach of this paper is divided into three phases: (a) automatically constructing a high-quality dataset of fix categories, based on existing datasets of flaky test cases and their fixes; (b) building a prediction model of fix categories, using the created dataset, to guide the repair of flaky tests; and (c) automatically generating the fixed version of a given flaky test using LLMs, with the help of our predicted fix categories.

 \subsection{Automatically Constructing a Dataset of Fix Categories}\label{subsec:categoriesFlakyTest}
 Having an appropriate and well-defined methodology for constructing a high-quality dataset of fix categories is important in practice, as we expect that local prediction models will need to be built in most contexts based on locally collected data. 
 
In a nutshell, our approach starts with an existing set of fix categories and then extends it to a more comprehensive and practical set, and finally devise heuristics to automatically label fixes in existing flaky test datasets.

To the best of our knowledge, the categories of flaky test fixes presented by Parry et al.~\cite{parry2022developer}, as shown in Table~\ref{tab:CausesAndRepair}, represent the largest and most recent classification in the literature. These fix categories were manually assessed by the authors by analyzing code diffs, developer's comments, and any linked issues of 75 flakiness-repairing commits that originated from 31 open-source Python projects \cite{parry2022developer}.

Their categorization procedure accepts multiple sources of information (i.e., fixes but also comments and linked issues) to link a flaky test with a fix category. This information might not be entirely available for many flaky tests in a project. Therefore, following that procedure, not all fixes can be labeled with a category if we have only access to the flaky tests and their fixes, which is the case for many datasets.

In this paper, we aim to categorize fixes and train a prediction model whose objective is to assign a fix category to a new flaky test, without any other other information than the test case's code. In other words, the goal is to guide testers (or automated repair tools) in applying the right fix according to categories of fixes that are predictable based on the code of the test cases alone. 
While not all flaky tests can be resolved solely by addressing the test code, a subset can indeed be fixed in this manner. Automating these fixes can still save time and resources. Hence, our dataset creation is targeting these specific fixes. 
% Not all flaky tests can be fixed by only considering the test code, but there is nevertheless a subset that can be fixed as such, hence saving time and resources if these fixes are automated. Thus we focus on such fixes while creating the dataset.
To achieve that, we analyzed flaky test cases and their fixes, to extract ground truth labels. Inspired by Parry et al.'s work \cite{parry2022developer}, we analyzed the International Dataset of Flaky Tests (IDoFT)\footnote{\url{https://mir.cs.illinois.edu/flakytests}}, which comprises flaky test cases along with their fixes from 123 open-source Java projects. 
% This dataset has been used by several previous studies on flaky test case prediction~\cite{wei2021probabilistic,lam2019idflakies,lam2020large,lam2020understanding,lam2020dependent,shi2019ifixflakies},\cite{fatima2022flakify}. This dataset is a public dataset and hence is updated regularly. We made sure to use the latest version of this dataset. 

This dataset, utilized in various prior studies on flaky test case prediction~\cite{wei2021probabilistic,lam2019idflakies,lam2020large,lam2020understanding,lam2020dependent,shi2019ifixflakies},\cite{fatima2022flakify}, is a publicly accessible dataset regularly receiving updates. We ensured our usage involved the most recent version of this dataset.

%We performed Inter-rater reliability analysis \cite{b10},\cite{b11}, which is commonly used for labeling or annotating software code. 
To assign labels to IDoFT dataset with fix categories, two researchers manually and independently analyzed 100 randomly selected flaky tests and their corresponding fixes to map them to one or more categories from those proposed by Parry et al. (see Table~\ref{tab:CausesAndRepair}) or a new category when the fix did not match any of them. 
The new categories should be predictable by only looking at the flaky tests and yet be useful enough to help developers and testers fix them.
Differences in categories and definitions between the researchers were resolved through multiple meetings.
We use Cohen’s Kappa~\cite{mchugh2012interrater} to measure the initial level of agreement between the two researchers. This is briefly discussed insection~\ref{Validation}. There was no case where a third opinion was required to resolve a conflict. 

Some of the fix categories proposed by Parry et al. are also renamed in Table~\ref{tab:CausesAndRepair} for better understandability. For example, the term \textit{Reduce Scope} (one of the Parry et al.'s categories) lacks clarity in indicating the exact code section that requires modification. By renaming \textit{Reduce Scope} to \textit{Change Condition}, developers and testers are more explicitly instructed to modify conditional statements. 
% For example ~\textit{Reduce Scope} is renamed to ~\textit{Change Condition} to exactly point it to the developers that there is need to change in conditional statements however ~\textit{Reduce Scope} does not exactly show which part of the code needs to be changed very clearly .
The process resulted in 13 different fix categories for flaky tests, as shown in Table ~\ref{tab:FlakyFixCategories}. 
% Note that a single test may belong to more than one fix category.

The final step in this phase was to create a set of heuristics to automatically label each fix from our dataset to one of these 13 categories.~Each heuristic is a search rule (implemented as a Python script) that relies on keywords from the~\textit{diff} of a flaky test case and its fixed version.
% Lastly, we manually verified the automated labeling outputs to ensure the correctness of the heuristics for future datasets.
Subsequently, we manually verified the outputs of the automated labeling process to ensure the accuracy of the heuristics for future datasets.~The complete labeling process is shown in Figure~\ref{fig:Labelling_process}.
Additionally, after this final step, the remaining 462 tests from our dataset were automatically labeled using the developed heuristics. Note that a single test may be labeled with multiple fix categories, as illustrated in Table ~\ref{tab:FlakyFixCategories}.

In the following, we briefly explain each category's heuristic along with few examples from the IDoFT dataset.
\begin{table}[!htbp]
    \centering
    \setlength{\arrayrulewidth}{0.8pt} % Adjust the padding here
    \resizebox{0.46\textwidth}{!}{
    \begin{tabular}{|c|c|c|c|c|c|c|c|c|c|c|}
        \hline
        \textbf{Cause} & \rotatebox[origin=c]{90}{ \textbf{Add Mock}} & \rotatebox[origin=c]{90}{ \textbf{Add/Adjust Wait}} & \rotatebox[origin=c]{90}{ \textbf{Guarantee Order}} & \rotatebox[origin=c]{90}{ \textbf{Isolate State}} & \rotatebox[origin=c]{90}{ \textbf{Manage Resource}} & \rotatebox[origin=c]{90}{ \textbf{Reduce Random}} & \rotatebox[origin=c]{90}{ \textbf{Reduce Scope}} & \rotatebox[origin=c]{90}{ \textbf{Widen Assertions}} & \rotatebox[origin=c]{90}{ \textbf{Miscellaneous}}& \rotatebox[origin=c]{90}{ \textbf{Total}}\\
        \hline
        \text{Async. Wait} & 1 & 6 & - & - & - & - & - & 2 & - & 9 \\
        \hline
        \text{Concurrency} & - & 2 & - & - & 2 & - & - & 2 & 2 & 8 \\
        \hline
        \text{Floating Point} & - & - & - & - & - & - & - & 3 & - & 3 \\
        \hline
        $\mathrm{I/O}$ & - & - & - & - & - & - & - & - & - & - \\
        \hline
        \text{Network} & 3 & 3 & - & - & 1 & - & - & - & 1 & 8 \\
        \hline
        \text{Order Dependency} & - & - & - & 2 & - & - & 1 & - & - & 3 \\
        \hline
        \text{Randomness} & - & - & - & - & - & 6 & - & 4 & 1 & 11 \\
        \hline
        \text{Resource Leak} & - & - & - & - & 2 & - & 1 & 1 & - & 4 \\
        \hline
        \text{Time} & 5 & - & - & - & - & - & 1 & 1 & 2 & 9 \\
        \hline
        \text{Unordered Coll.} & - & - & 3 & - & - & - & - & - & - & 3 \\
        \hline
        \text{Miscellaneous} & 2 & - & 1 & - & 1 & - & - & 6 & 7 & 17 \\
        \hline
        \text{Total} & 11 & 11 & 4 & 2 & 6 & 6 & 3 & 19 & 13 & 75 \\
        \hline
    \end{tabular}
    }

    \caption{ Each flaky test is assigned a category describing the cause of
flakiness (rows) and another describing the repair applied by developers
(columns). From Parry et al.~\cite{parry2022developer}}
    \label{tab:CausesAndRepair}
\end{table}

%Table 2: Fix Categories Table: 

\begin{table}[!htpb]
    \setlength{\tabcolsep}{0.25em}
    \renewcommand{\arraystretch}{1.5}
    \centering
    \resizebox{0.38\textwidth}{!}{
        \begin{tabular}{|c|c|}
            \hline
            \textbf{Flaky Test Fix Category} & \textbf{Number of Flaky Tests} \\
            \hline
            \highlightrow{Change Assertion} & \highlightrow{197} \\
            \highlightrow{Change Condition} & \highlightrow{121} \\
            \highlightrow{Reset Variable} & \highlightrow{104} \\
            \highlightrow{Reorder Data} & \highlightrow{92} \\
            \highlightrow{Change Data Structure} & \highlightrow{91} \\
            \highlightrow{Handle Exception} & \highlightrow{52} \\
             \highlightrow{Miscellaneous} &  \highlightrow{50}\\
             \highlightrow{Change Data Format}&  \highlightrow{46}\\
             \highlightrow{Reorder Parameters} &  \highlightrow{37} \\
            Call Static Method & 9 \\
            String Matching & 6 \\
            Change Timezone & 3 \\
            Handle Timeout & 2 \\
            \hline
        \end{tabular}
    }\vspace{0.2 cm}  % Adjust the space between table and caption as needed
    \caption{Count of flaky tests within each fix category, using automated labeling. Categories with over 30 instances are highlighted. }
    \label{tab:FlakyFixCategories}
\end{table}

For the \textbf{Change Assertion} category, the heuristic looks for the assert keyword and its replacements within the fixed flaky test code. For instance, if \emph{AssertEquals}, \emph{AssertThat}, or \emph{AssertTrue} appear in deleted lines and \emph{AssertJsonStringEquals}, \emph{assertJSONEqual}, \emph{AssertJsonEqualsNonStrict}, or \emph{JsonAssert.AssertEquals} are present in the added lines, it shows an assertion change to address flakiness.

A common instance of flakiness arises due to the use of an incorrect assertion type. For example,~\emph{assertJSONEqual} should be used instead of~\emph{AssertThat} in an assert function that retrieves a JSON object from other functions in the test code and compares it with a hard-coded JSON String to ensure the deterministic ordering of the retrieved JSON object.  An example illustrating the appropriate assertion type usage can be found online~\footnote{\url{https://github.com/vipshop/vjtools/pull/168}}.

Changing data structures can be an effective technique to improve test reliability and reduce flakiness~\cite{meszaros2007xunit,osherove2013art}. For example, the data structures \emph{Hash} \emph{Map}, \emph{Set}, \emph{HashMap}, and \emph{HashSet} are unordered collections. Their use in the test code can lead to test flakiness if the test relies on the order of elements in the structure. In such cases, these data structures should be replaced with \emph{LinkedHash}, \emph{LinkedMap}, \emph{LinkedSet}, \emph{LinkedHashMap} and \emph{LinkedHashSet}, respectively, as they maintain the order in which their elements are stored, regardless of how many times a test is executed. This makes the test more reliable and less prone to flakiness.
If \emph{Hash}, \emph{Map}, \emph{Set}, \emph{HashMap}, and \emph{HashSet} keywords are replaced in the test code 'diff' with \emph{LinkedHash}, \emph{LinkedMap}, \emph{LinkedSet}, \emph{LinkedHashMap} and \emph{LinkedHashSet}, respectively,  then such test cases are labeled as ~\textbf{Change Data Structure}. An example of this type of fix can be found online~\footnote{\url{https://github.com/biojava/biojava/pull/897}}. 

Exception handling is usually an important tool to reduce unwanted behavior due to wrong inputs or scenarios and thus decrease flakiness. If there are \emph{Try Catch} statements added or removed or changes occur in the type of the Exception used in the test case, the test case is labeled as~\textbf{Handle Exception}.
Handling exceptions, similar to other fix categories, can be combined with different fix strategies to entirely eliminate the root cause of flakiness within a test case. Examples of this can be seen online~\footnote{ \href{https://github.com/apache/incubator-hugegraph-toolchain/pull/398/files}{https://github.com/apache/incubator/pull/398}}\textsuperscript{,}
\footnote{\url{https://github.com/vipshop/vjtools/pull/168/files}}where flakiness is fixed in the test codes after integrating exception handling techniques alongside alterations in assertion types.

A flaky test is labeled with the~\textbf{Reset Variable} category if one of the keywords \emph{reset}, \emph{clear}, \emph{remove} or \emph{purge} is found in the ~\textit{diff}. These are useful commands to reset the state of the system and clear the memory at the end of the test cases. They put the system back to its initial testing state, which is required for many tests to avoid flakiness. An example of this type of fix can be found online~\footnote{\url{https://github.com/apache/nifi/pull/6754}}.

Input formatting mismatches can be a cause of flakiness. Oftentimes, a fix is simply changing the input variable's format to the required one. If there is change in the format of the string or numeric data that is used in a function of the test code, the test case is labeled as~\textbf{Change Data Format}. 

If conditional statements are added or removed in the test code, for example replacing the \emph{containsexactly} keyword with the \emph{containsexactlyinanyorder} keyword, the test case is labeled with the \textbf{Change Condition} category. An example of this fix is shown in this pull request~\footnote{\url{https://github.com/apache/servicecomb-java-chassis/pull/3393}}.
If there is any of the keywords~\emph{sortfield},~\emph{sort\_properties},~\emph{sorted},~\emph{order by} or ~\emph{sort} in the fixed test code, which is used to sort or change the order of data, then the test is labeled as \textbf{Reorder Data}.  An example of this type of fix can be found~online\footnote{\url{https://github.com/cerner/common-kafka/pull/50}}.

If there is a change in the order of the function parameters, then the \textbf{Reorder Parameters} label is used.  An example of this type of fix can be found online~\footnote{\url{https://github.com/apache/nifi/pull/6709/files}}.

Tests may rely on specific dates or times, which can be affected by time zone differences, leading to inconsistent test results. Setting a specific time zone ensures consistent results, increasing test reliability and reducing flakiness~\cite{luo2014empirical}~\cite{zheng2021research}. If there is a change in the assigned value to the~\emph{timezone} keyword, 
%for example changing timezone's value from US/Pacific to Europe/London in the fixed test code,
then such tests are labeled as \textbf{Change Time Zone}.

If the ~\emph{Timeout} keyword is added to the test code, which ensures that tests are completed within a reasonable time frame, thus reducing the likelihood of flakiness, then these test cases are labeled as~\textbf{Handle Timeout}. This fix has been suggested in the work of Pei et al.~\cite{pei2023traf} to fix flakiness related to Async Wait.

Static methods can help ensure consistent test behavior across multiple runs by encapsulating complex scenarios or behavior simulations, such as network failures or time delays. If there is a static method call added to the test code, then these tests are labeled as \textbf{Call Static Method}. 
If two strings are matched in a given test to make sure they are in the correct order for maintaining determinism by using keywords like~\emph{'match.that(s).isequalto} or~\emph{.matches} then these tests are marked as \textbf{String Matching}. An example of this type of fix can be found online~\footnote{\url{https://github.com/immutables/immutables/pull/1350}}.
Lastly, if there is some change in the added or deleted lines of code that do not belong to any of the above categories, the test cases are labeled as \textbf{Miscellaneous}. Examples of such fixes that deviate from established patterns in other fix categories can be located online~
\footnote{\href{https://github.com/SAP/emobility-smart-charging/pull/42/files}{https://github.com/emobility-smart-charging/pull/42}}\textsuperscript{,}\footnote{\url{https://github.com/sofastack/sofa-boot/pull/1124}}. 
 
\subsection{Flaky Test Case Fix Category Prediction with Code Models}
\label{Prediction with Code Models}
In this section, we describe how we use different language models for predicting the fix category of a flaky test. This prediction may be not only useful to help testers identify the required fix but can also guide any automated repair process. 
% Though our prediction objectives are different, one of our prediction model's overall approach is similar to that of FlakyCat~\cite{akli2023flakycat}. 

We use two different code models CodeBERT and UniXcoder as our prediction models, respectively. In general, language models are pre-trained using self-supervised learning, on a large corpus of unlabelled data. They are then further fine-tuned using a specific, labeled dataset for different Natural Language Processing (NLP) tasks such as text classification, relation extraction, or next-sentence prediction~\cite{feng2020codebert}. Code models are those language models that are specifically pre-trained on programming languages. 
CodeBERT and UniXCoder have been pre-trained using a diverse corpus encompassing code from six programming languages: Java, Python, JavaScript, PHP, Ruby, and Go. Both models excel in code classification tasks when compared to other pre-trained code models~\cite{niu2023empirical,fatima2022flakify}. These models comprehend both the syntax and semantics of various programming languages. We individually fine-tune these code models on our dataset of flaky tests to learn the underlying patterns of test code flakiness. During inference, these fine-tuned models classify flaky tests based on what these code models learned during training (fine-tuning). 
We choose not to fine-tune Large Language Models (LLMs) like GPT or CodeLlama for fix category prediction due to the high costs associated with their fine-tuning. In addition, as we will see in the results section the smaller code models will already provide very high accuracy which leaves no reason for us to switch to much more complex and heavier models. This process demands extensive data and substantial computational resources~\cite{zhao2023survey}. Reported studies show that smaller code models work well for classifying code~\cite{niu2023empirical} and are more cost-effective. Their efficient classification abilities, along with reduced time and computational expenses for refinement, make them more practical for integration into solutions for large-scale industrial systems.

For fine-tuning CodeBERT and UniXcoder to predict flaky test fix categories, we use two different techniques. In the first one, we augment the models with FSL using a Siamese Network to deal with the usually small datasets of flaky tests that are labeled in a typical development context. In contrast to FlakyCat, that also uses FSL~\cite{akli2023flakycat}, we train our model on the dataset created following the procedure presented in section \ref{subsec:categoriesFlakyTest}, where each data item is a pair of flaky test and its fix category. Note that we do not use information from the actual fix during inference time, since our goal is to predict the fix category for a given flaky test which has not been repaired yet. Thus, in practice, only the test code is available to the model. 
In the second technique, we fine-tuned the models using a simple Feed Forward Neural Network (FNN) as done in earlier work for flaky test classification~\cite{fatima2022flakify}.

Furthermore, note that we do not build a multi-class prediction model to deal with multiple fix categories. Instead, we build multiple binary classification models to estimate the probability of whether a test belongs to each category. This is because flaky tests may belong to multiple fix categories. With multiple binary classifications, each fix category is treated separately, allowing the model to learn the features that are most relevant for that fix category.~Since our dataset is relatively small, it is not computationally expensive to build these multiple classifiers for each category. However, as our dataset expands to include more samples, we may consider transitioning to a multi-class classifier for efficiency and scalability purposes in the future.
\subsubsection{Converting Test Code to Vector Representation}
Prediction using language models starts with encoding inputs into a vector space.  In our case, we embed test cases into a suitable vector representation using  CodeBERT and UniXcoder, respectively. 
% as shown in the Figure~\ref{fig:FSL_SiameseArchitecture_new}.

Both CodeBERT and UniXcoder take the test case code as input and convert it into a sequence of tokens. These tokens are then transformed into integer vector representations (embeddings).
The test case code is tokenized using 
Byte$-$Pair Encoding (BPE)~\cite{sennrich2015neural}, which follows a segmentation algorithm that splits words into a sequence of sub-words.

Each token is then mapped to an index, based on the position and context of each word in a given input. These indices are then passed as input to the CodeBERT and UniXcoder models to generate a vector representation for each token as output. 
Finally, for a test case, all vectors are aggregated to generate a vector called~\texttt{[CLS]}~\cite{fatima2022flakify}. 

\subsubsection{Fine-tuning Code Models}
As discussed earlier, we have fine-tuned CodeBERT and UniXcoder independently with two different techniques. Given our limited labeled training data, which is a common situation in our application context, few shot learning approaches~\cite{wang2020generalizing}, such as the Siamese Network architecture~\cite{he2018twofold}, are adequate candidate solutions.
%what is a Siamese Network? 
A Siamese Network trained with a specific loss function (\textit{Triplet Loss Function}~\cite{schroff2015facenet}) is designed to determine the similarity between inputs by learning an embedding space~\cite{neculoiu2016learning}—a high-dimensional vector space where inputs are represented as embeddings or vectors. In this space, similar inputs are positioned closer together, while dissimilar inputs are placed farther apart.
Similar to existing work~\cite{akli2023flakycat}, 
we constructed a Siamese Network~\cite{koch2015siamese,mueller2016siamese,akli2023flakycat} with three identical sub-networks. Each sub-network is based on either CodeBERT or UniXcoder, essentially using the same model (weights and network parameters) three times within the network structure.
The inputs to this Siamese Network are 
\textit{anchor}, \textit{positive} and \textit{negative} tests, which are transformed by each sub-network of the Siamese Network into high-dimensional embeddings~\cite{neculoiu2016learning}.
%How does it work?
The anchor and positive tests, which belong to the same class, are projected into the embedding space such that their distance is minimized, while the distance between the anchor and the negative test, belonging to a different class, is maximized. This process ensures that similar inputs are grouped together in the vector space, facilitating tasks such as similarity comparison or classification based on learned similarities.~During training, the triplet loss function fine tunes the embeddings by computing similarity scores, encouraging the network to improve its ability to distinguish flaky tests belonging to different categories. 
%What we do?
In our approach, given a category (e.g., \textit{Change Assertion}), the anchor is a flaky test in the training dataset that belongs to the positive class of that category, e.g., it is labeled as \textit{Change Assertion}. A positive sample is another flaky test in the training dataset that belongs to the same class as the anchor (e.g., labeled again as \textit{Change Assertion}) and is considered similar to the anchor. The negative sample is another flaky test from the dataset that belongs to a different class than that of the anchor (e.g., NOT labeled as \textit{Change Assertion}) and is considered dissimilar to the anchor. 
More details about the procedure to create these samples in our experiment will be provided in Section \ref{Validation}. 

% By training the Siamese Network with a specific loss function (\textit{Triplet Loss Function}~\cite{schroff2015facenet}), the model learns to generate~\texttt{[CLS]} embeddings for each anchor test case in a way that it captures the similarity between the anchor and positive examples while minimizing the similarity between the anchor and negative examples.

\begin{figure*}[hbt!]
    \centering
    \hspace*{-1.1cm}
    \includegraphics[width=1.12\linewidth]{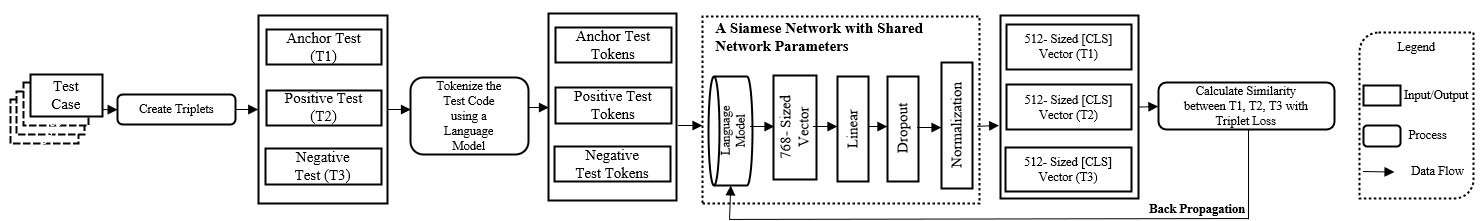}
    \caption{Training of a Siamese network with a shared language model and network parameters.}
    \label{fig:FSL_SiameseArchitecture_new}
    \vspace{-3pt}
\end{figure*}

We use the same architecture as Akli et al~\cite{akli2023flakycat} for each sub-network. First, it includes a linear layer of 512 neurons. To prevent model over-fitting, we then add a dropout layer to randomly eliminate some neurons from the network, by resetting their weights to zero during the training phase~\cite{el2021bert}. 
Lastly, we add a normalization layer before each sub-network that generates a~\texttt{[CLS]} vector representation for the input test code of that sub-network. Figure~\ref{fig:FSL_SiameseArchitecture_new} depicts the architecture of a Siamese Network where anchor, positive and negative tests share a language model and other network parameters to update the model's weights based on the similarity of these tests. 
For training both CodeBERT and UniXcoder, we use the AdamW optimizer~\cite{yao2021adahessian} with a learning rate of $10^{-5}$ and a batch size of two, due to available memory (8 gigabytes).

To evaluate the final trained model, following FSL, we create a support set of examples, for each prediction task. As a reminder, each prediction task here is a binary prediction of flaky tests indicating whether they belong to a fix category (positive class) or not (negative class).  The support set includes a small number of examples for each class, from the training dataset. For every flaky test case in the support set of each class (positive and negative), we first calculate their ~\texttt{[CLS]} vector embeddings. We then calculate the mean of all test case vectors in that class, to represent the centroid of that support set class.

\begin{figure*}[hbt!]
    \centering
    \hspace*{-0.5cm}
    \includegraphics[width=1.05\linewidth]{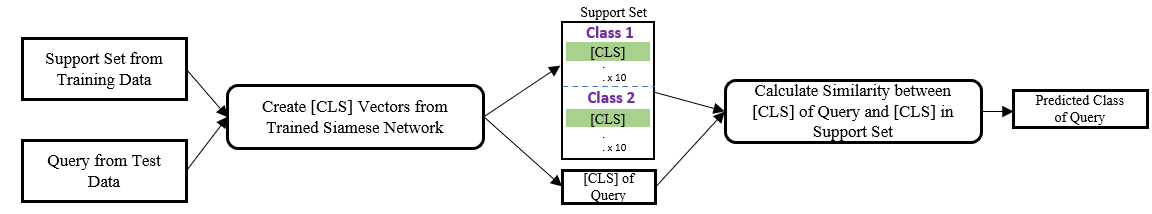}
    \caption{Predicting the class of query test case using few shot learning and a trained Siamese network.}
    \label{fig:full_architecture}
    \vspace{-3pt}
\end{figure*}

Each flaky test in the test data is referred to as a ~\textit{query}. To infer the class of a query, from the test data, we first calculate the vector representation of the query using the trained model. Then using the cosine similarity of the query and the centroid of the classes, we determine to which class (positive or negative), the query belongs, i.e., the one with the highest similarity.
Figure~\ref{fig:full_architecture} shows the process of predicting the class of a query using FSL.
%Explain here the second method of fine-tuning

\begin{figure*}[hbt!]
    \centering
    \includegraphics[width=0.8\linewidth]{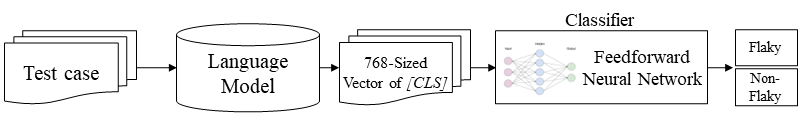}
    \caption{Fine-tuning language model for classifying tests into different fix categories.~\cite{fatima2022flakify}}
    \label{fig:FNN_FineTuning}
    \vspace{-3pt}
\end{figure*}

Regarding our second method for fine-tuning CodeBERT and UniXcoder, we train a Feedforward Neural Network (FNN)  to perform binary classification for each fix category. This process is depicted in Figure~\ref{fig:FNN_FineTuning} and was used by Flakify, which relied on  CodeBERT~\cite{fatima2022flakify}. 
The output of the language model, i.e., the\texttt{[CLS]} token which is the aggregated vector representation token of all input tokens, is then fed as input to a trained FNN to classify tests into the fix categories. The FNN contains an \textit{input} layer of 768 neurons, a \textit{hidden} layer of 512 neurons, and an \textit{output} layer with two neurons. Like previous work~\cite{fatima2022flakify}, we used ReLU~\cite{agarap2018deep} as an activation function, and a \textit{dropout} layer~\cite{yao2021adahessian}. Lastly, we used the \textit{Softmax} function to compute the probability for a flaky test to belong to a given fix category. 

\subsection{Repairing Flaky Tests Using GPT and our Proposed Fix Categories}\label{Repairing_Flaky_Tests_Using GPT_and_Proposed_Fix_Categories}
Large Language Models (LLMs) such as GPT have been successfully applied to many automated software engineering tasks in the past~\cite{zheng2023survey,fan2023large}, including automated program repair~\cite{pearce2023examining,white2023chatgpt,zhang2023critical} and~\cite{fu2023chatgpt}.
Therefore, as part of our study, we are interested to see (a) how effectively an off-the-shelf GPT can generate fixes for our flaky tests and (b) how our classification of fixes can potentially help GPT to better fix flakiness. 

We use GPT-3.5 Turbo, with 175 billion parameters and a context window of 16,385 tokens, designed to accommodate long inputs (in our case full test cases). This GPT version was released on November 06, 2023 and has shown superior performance for diverse code-related tasks compared to prior versions. As per OpenAI's official website~\cite{openaiplatform}, this model has been trained on data up to September 2021. 

GPT-4, which was unveiled early 2023, features 1.7 trillion parameters, significantly surpassing GPT-3.5 in performance. However, the latest GPT-4's training dataset extends up to December 2023. Further, there is another version of GPT-4 that is trained up to April 2023.  Given that a majority of flaky tests in our dataset were posted on the public repository before these two dates, we chose not to use GPT-4 as it could lead to  data leaks. Though there are more recent test case fixes in our dataset, most of them are not yet accepted changes and we only consider fixes that are approved by the developers. 

%Our study utilizes open source flaky test data from the IDoFT repository, a comprehensive compilation encompassing flaky tests from recent research endeavors. Given that many pull requests remain pending and considering the delay in their acceptance due to the repository's open-source nature, there's a scarcity of cases accepted post-April 2023. Consequently, to comprehensively evaluate GPT's performance across a broader spectrum of tests, we rely on GPT-3.5 for tests added with accepted fixes post-September 2021. Our goal is not merely to generate fixes for flaky tests using GPT but to scrutinize how predicted fix categories can augment the generation process, achievable even with the latest version of GPT-3.5.

\subsubsection{GPT Prompt for Fixing Flaky Tests}
To evaluate the impact of predicted fix categories on GPT's repair capability, we incorporate the predicted category alongside the flaky test code as input query to GPT (the prompt). We use two different methods to build the prompt. The first simply provides flaky test code with its corresponding fix category label, while the second employs in-context learning. Influenced by recent research on prompt optimization for GPT~\cite{liu2021makes}, in-context learning provides multiple examples of inputs and their expected outputs, as the context. Since fine-tuning a LLM is expensive, with in-context learning, the LLM learns at inference time~\cite{workrethinking} on how to optimize its output for a particular problem domain (characterized by the examples in the context). 

Section~\ref{prompt_design} shows our both prompt templates. In our current implementation of in-context learning, a prompt contains the flaky test requiring repair, its predicted category, and some sample flaky tests from the same category in the training set, accompanied by their respective fixes. This methodology enriches the model's output (the fix) by exposing it to prevalent resolution strategies applied to flaky tests within a specific category. Future investigations could delve into further optimizing these in-context learning prompts~\cite{qiao2022reasoning} to get the best results possible from GPT.
\begin{figure*}[hbt!]
    \centering
    \hspace*{-1.1cm}
    \includegraphics[width=1.1\linewidth]{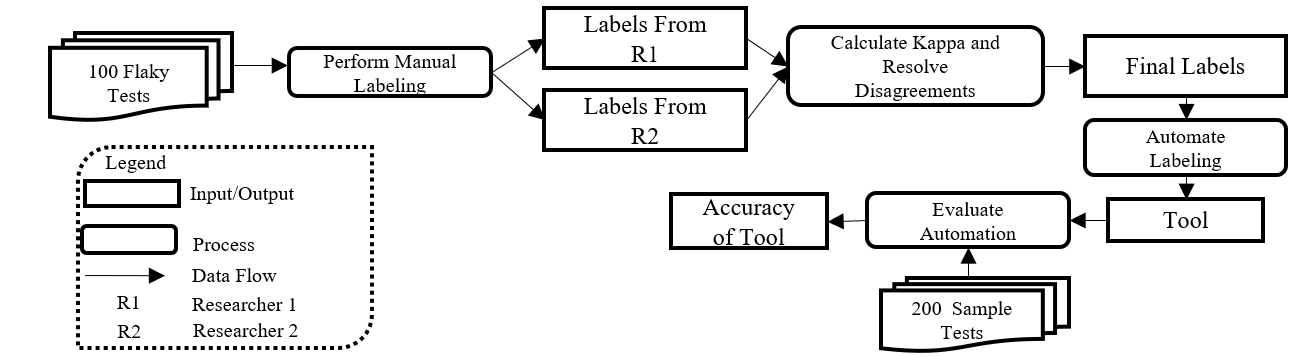}
    \caption{Automated labeling of flaky tests with their respective fix categories.}
    \label{fig:Labelling_process}
    \vspace{-3pt}
\end{figure*}

\newcommand{\flakyTestResolution}{%
    \begin{tikzpicture}[node distance=0.45cm] % adjust vertical distance
        \node [draw, rectangle, fill=white, text width=7.0cm, align=left, font=\ttfamily, outer sep=0pt, inner sep=5pt] (rect1) {
            \faPencil* \hspace{0.1cm}  \textbf{Input(Prompt):} \\
                         This test case is Flaky: 
            \textbf{[Flaky Code]}\\
            This test can be fixed by changing the following information in the code:
            \textbf{[Fix Category Label]}
            \\
            Just Provide the full fixed code of this test case only without any other text description.
        };
        \node [below=of rect1, outer sep=0pt, inner sep=0pt] (img) {\includegraphics[width=4cm]{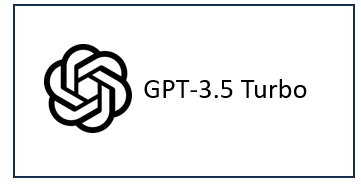}};
        \node [below=of img, draw, rectangle, fill=white, text width=7cm, align=left, font=\ttfamily, outer sep=0pt, inner sep=5pt] (rect2) {
            \faPencil* \hspace{0.1cm} \textbf{Output(Generated Output):} 
            \textbf{[Fixed Flaky Test Code]}
        };
        \draw[->, thick] (rect1.south) -- (img.north);
        \draw[->, thick] (img.south) -- (rect2.north);
    \end{tikzpicture}
}
\newcommand{\flakyTestResolutionincontextLearning}{%
    \begin{tikzpicture}[node distance=0.45cm] % adjust vertical distance
        \node [draw, rectangle, fill=white, text width=7.0cm, align=left, font=\ttfamily, outer sep=0pt, inner sep=5pt] (rect1) {
            \faPencil* \hspace{0.1cm}  \textbf{Input(Prompt):} \\
            This test case is Flaky: 
            \textbf{[Flaky Code]}\\
            This test can be fixed by changing the following information in the code:
            \textbf{[Fix Category Label]}
            \\
            Just Provide the full fixed code of this test case only without any other text description.
            Here are some Flaky tests examples, their fixes and fix category labels:
            \textbf{[Examples]}
        };
        \node [below=of rect1, outer sep=0pt, inner sep=0pt] (img) {\includegraphics[width=4cm]{Images/GPT3.5_logo.jpg}};
        \node [below=of img, draw, rectangle, fill=white, text width=7cm, align=left, font=\ttfamily, outer sep=0pt, inner sep=5pt] (rect2) {
            \faPencil* \hspace{0.1cm} \textbf{Output(Generated Output):} 
            \textbf{[Fixed Flaky Test Code]}
        };
        \draw[->, thick] (rect1.south) -- (img.north);
        \draw[->, thick] (img.south) -- (rect2.north);
    \end{tikzpicture}
}
\section{Validation} \label{Validation} 
The goal of this study is to predict fix categories for a flaky test to help generate a fully repaired version of the test. This section validates our approach (See Section \ref{Approach}) towards that goal by first introducing our research questions and explaining the data collection procedures. Then it discusses the experimental design and report the results. 

\subsection{Research Questions}
In this study, we address the following research questions: 
\paragraph*{\textbf{RQ1}} How accurate is the proposed automated flaky test fix category labeling procedure? \\
\textbf{Motivation}: Given that there is no existing dataset of flaky tests that is labeled by fix categories, in this RQ, we validate our proposed procedure to automatically label a flaky test given its fix. If accurately automated, such a procedure is quite useful since it can convert any flaky test dataset that includes fixes, to an appropriate training set for building prediction models, such as those used in RQ2. 
In practice, a customized training set (from local projects) is often necessary to capture domain knowledge, either to train new models from scratch or fine-tune pre-trained models, which further justifies the need for an automated labeling technique.
  
\paragraph*{\textbf{RQ2}}How effective are different language models, i.e., CodeBERT and UniXcoder with and without FSL, in predicting the fix category, given flaky test code? \\
\textbf{Motivation}: 
Most existing prediction models related to flaky tests try to predict flakiness. FlakyCat~\cite{akli2023flakycat} is, however, the only approach that classifies flaky tests according to their root cause using CodeBERT and FSL. Identifying the root cause alone, however, does not directly aid testers in resolving flakiness. For instance, one suggested root cause in FlakyCat~\cite{akli2023flakycat} is ~\textit{Unordered Collections}, highlighting that it causes flakiness. However, the proposed solution categories in ~\cite{parry2022developer} maps this cause to the ~\textit{Guarantee Order} fix category without providing concrete guidance on how to ensure such order.
In contrast, our defined fix categories offer a one-to-one mapping of specific fixes to individual test cases. For instance, concerning ~\textit{Unordered Collections} cause, Table~\ref{tab:FlakyFixCategories}~presents multiple fix categories such as ~\textit{Reorder Data}, ~\textit{Change data structure} or ~\textit{Reorder Parameters}. Then depending on the test code one of these concrete fixes is going to be predicted as the fix category.  This approach provides testers with more precise solutions rather than a broad label like ~\textit{Guarantee Order}.

This practical difference between a fix and a root-cause, led us to investigate alternative approaches for classifying the tests according to their required fix category, given that these categories are based on actual fix patterns.~In this RQ, we explore two code models as representatives of the state of the art. We also study the effect of FSL with each model, since FSL is usually recommended when the training data set is small, like in our case.
\paragraph*{\textbf{RQ3}}How effective is GPT-3.5 Turbo in repairing a given flaky test, with and without information regarding its fix category and fix examples?\\
\textbf{Motivation}: Given a fix category, the next question is to check to what extent having that information helps in fixing flakiness. To answer that, we focus on a fully automated flaky test repair approach and want to assess whether knowing the flaky test category help obtain better repairs. To automate the repair process, we chose to rely on LLMs that are pre-trained on an extensive corpus coming from various resources, enabling them to generate complete code without requiring fine-tuning.\\
The recent development of LLMs with 175 billion parameters for GPT-3.5 Turbo and 1.7 trillion parameters for GPT-4~\cite{openaiplatform}, trained over a dataset size of 45TB~\cite{brown2020language} offers significant potential for automating tasks related to source code, including automated program repair~\cite{fu2023chatgpt}. Thus RQ3 investigates whether the fix categories predicted in RQ2 enhance the performance of GPT models in repairing flaky test cases and to what extent we can achieve correct repairs. Additionally, we use in-context learning to understand how the GPT model's ability to fix issues improves when given examples of a specific type of flaky fix alongside our suggested fix category label.

\subsection{Dataset and Data Augmentation:}
To create a labeled dataset of flaky test fix categories for our study, we first analyzed various datasets of flaky tests that include fixes. Then we picked the most suited dataset and augmented it with our automatically generated labels.

\subsubsection{Existing Datasets}
There is no publicly available dataset of flaky tests that have been labeled based on the nature of their fix. Most existing datasets for flaky tests, such as FlakeFlagger \cite{alshammari2021flakeflagger}, simply contain binary labels indicating their flakiness. The dataset presented in the most related work, FlakyCat~\cite{akli2023flakycat}, assigns a label to each flaky test based on the category of flakiness causes, not their fixes. 
Therefore, we needed to build our own labeled dataset out of an unlabeled one that consists of both flaky tests and their fixes.
We also required a dataset with a sufficient variety of flaky test fixes.
Previous research on flaky test repair \cite{dutta2021flex} has relied on datasets of limited size and scope. 
For instance, the Flex~\cite{dutta2021flex} dataset comprises only 35 flaky tests. Eck et al. proposed a study based on the Mozilla database of flaky tests\cite{eck2019understanding} where they analyzed 200 flaky tests repaired by developers. 
Another example is the ShowFlakes dataset~\cite{parry2022developer} that includes 75 commits from 31 Python projects. However, it only includes 20 flaky tests where flakiness can be removed by only changing the test code.  
The iPFlakies and iFixFlakies datasets \cite{wang2022ipflakies, shi2019ifixflakies} are larger in size compared to previous datasets, but the root cause of flakiness in these datasets is related to the order in which the tests are executed, not the test code logic.

% The largest publicly available and relevant dataset is the International Dataset of Flaky Tests (IDoFT)\footnote{\url{https://mir.cs.illinois.edu/flakytests}}, an open-source dataset that is continuously updated. For this research, we collected a Java dataset of Flaky tests from IDoFT until October 23, 2023. This dataset notably contains 5500 tests, from which 128 duplicate tests were removed. Out of the remaining 5372 tests, 562 were identified as being fixed solely by altering the test code, thus aligning with the focus of our research. While the number of such tests where test code is fixed  are very less, however objective is to help testers who
% often do not have access to the production code and to only focus on the type of flaky fixes where test code is altered. 
The largest publicly available and relevant dataset is the International Dataset of Flaky Tests (IDoFT)\footnote{\url{https://mir.cs.illinois.edu/flakytests}}, an open-source dataset that is continuously updated. In this research, we collected a Java dataset of flaky tests from IDoFT until October 23, 2023. 
The dataset originally contained 5,500 tests, including 128 duplicates. After removing duplicates, we were left with 5,372 tests. We then filtered this set to include only those flaky tests for which the fixes were accepted, meaning that the pull requests (PRs) addressing these flaky tests on GitHub were approved by the original developers, thus resulting in 788 tests.
From this subset of flaky tests, we analyzed the diff of the original flaky tests and their developer repairs. We removed 226 tests that had repairs exclusively in the code under test (CUT) with no changes to the test code, leaving 562 tests that were fixed by modifying the test code.
As explained in the Introduction section as well, the flaky tests that can be fixed in a black-box manner are in the minority, but they are still important since testers and testing tools can fix them even without any access to production code, which is typical for some QA teams.
The remaining tests were resolved through test configuration adjustments, changes in test case execution order, fixes implemented in the production code, or are pending in open pull requests (awaiting approval, rejection, or being deliberately skipped by the original developers).
More details about this can be found online on the IDoft Repository. 
% This dataset comprises 5500 tests, including 128 duplicate tests. After excluding duplicates, the remaining 5372 tests were analyzed, revealing that 562 tests were fixed solely by modifying the test code and are therefore aligned with our research objectives. 
% As explained in the Introduction section as well, the flaky tests that can be fixed in a black-box manner are in the minority, but they are still important since testers and testing tools can fix them even without any access to production code, which is typical for some QA teams.
%Though the number of tests where the fix is confined to test code alterations represents a minority, we aim here to assist testers who frequently lack access to production code and need to repair the test code itself.
% We ensured that these fixes were accepted, meaning that the pull requests aimed at resolving these flaky tests on GitHub were approved by the original developers.
% The remaining tests were resolved through modifications in the test configuration, alterations in the order of test case executions, fixes in the production code, or their pull requests are not accepted and remain open, awaiting acceptance or these fixes are rejected or skipped by the original developers.  
% The remaining tests were resolved through test configuration adjustments changes in test case execution order, fixes implemented in the production code, or are pending in open pull requests (awaiting approval, rejection, or being deliberately skipped by the original developers).
% More details about this can be found online on the IDoft Repository. 
Consequently, our final dataset comprises 562 flaky tests and their corresponding resolutions from the IDoFT Java dataset. Notably, this dataset represents the largest available repository of flaky tests that are fixable through modifications of the test code. These tests originate from 123 distinct projects, offering a broad spectrum of test cases, which greatly benefits the generalizability of our predictive models.
\subsubsection{Labeled Dataset of Flaky Test Fix Categories}
To create the labeled dataset out of the IDoFT dataset, we followed the procedure discussed in Section \ref{Approach} and labeled each flaky test automatically.
Our dataset is publicly available in our replication package~\footnote{\url{https://figshare.com/s/47f0fb6207ac3f9e2351}}. 
The resulting labeled dataset is unbalanced with respect to the distribution of tests across fix categories.
Though, in our prediction phase, to handle the lack of data, we apply Few-Shot learning, to effectively predict the fix categories will require a minimum number of test cases per category. 
Therefore, we selected the categories with at least 30 tests as our training set. 
As a result, our labeled dataset consists of the top nine fix categories outlined in Table~\ref{tab:FlakyFixCategories} for training our model.
Since a flaky test may be associated with multiple fix categories, we have transformed our multi-label dataset into nine distinct bi-labeled sets, one per category. 
In each of the nine sets, the positive class includes tests that belong to the corresponding fix category and the negative class includes all the other tests.
This allows us to treat the problem of predicting each label (category) as a separate binary classification problem. 

\subsubsection{Data Augmentation for FSL}
We investigated code models with and without using FSL. Without FSL, we fine-tuned the code models with an FNN as discussed in Section~\ref{Approach}. For this, we used random oversampling~\cite{branco2016survey},\cite{fatima2022flakify}, which adds random copies of the minority class to our training dataset, which otherwise would be imbalanced (fewer positive examples).

In our fine-tuning approach, inspired by FlakyCat~\cite{akli2023flakycat}, we addressed limited training data by employing data augmentation techniques~\cite{shorten2021text,verdonck2021special} along with a Siamese Network, triplet loss function, and FSL. 
Specifically, we synthesize additional valid triplets for training, by creating $<$anchor, positive, and negative$>$ tests for each flaky test in the training data, thus increasing the training samples. For example, if the original training dataset has 20 samples, data augmentation will potentially create 20 new positive samples and 20 new negative samples, resulting in a total of 60 training samples.
For each triplet per category, the positive test must belong to the positive class (same as the anchor test), while the negative test must belong to the negative class.
To synthesize additional positive and negative tests, we mutate the test code by replacing the test names and changing the values of variables (integers, strings, and Boolean values) with randomly generated values. 
For example, for a test case in the training set named \textit{postFeedSerializationTest}, we create a new test case (and label it with the same class label) and call it \textit{postFeedSerializationTest\_xacu4}. Then we change the value of an integer variable 'access point' from '1' to '8967', randomly. Also, the string value is randomly replaced from 'AMAZON-US-HQ' to 'AMAZON-"RZjqJ"-HQ'. 

We automatically verify the created pairs, to ensure no duplicate tests are generated and to retain all flakiness-related statements unchanged.
% We make sure there is no duplicate test created in this process and retain all flakiness-related statements in the code.  
We thus produce enough samples to form the triplets (a set of anchor, positive and negative tests) required for this fine-tuning procedure. 

\subsubsection{Data Selection for Evaluation with GPT Model}
In RQ3, we selected 562 tests to assess how well the GPT model deal with our flaky tests with and without fix category. Since the GPT 3.5 model was trained on collected data up to September 2021, we only included tests that were fixed or added after that date.~Including older tests could be considered data leakage, since these tests are in the public domain (GitHub) and there is a high chance that they are in the model's training set. This exclusion criterion resulted in 181 tests from 61 distinct projects which included at least one test from each of the nine fix categories. 

\subsubsection{Data Selection for Execution of Repaired Flaky Tests}\label{ref:Dataselection_execution}
To gain further insights into the GPT-generated repaired flaky tests, we conducted executions of selected tests. Our dataset comprises 181 tests from 61 distinct projects. Among these projects, we selected projects where at least five flaky test cases were present, which resulted in nine projects. Out of these nine projects, the test cases of four projects could not be executed due to missing dependencies, resulting in build failures, or the unavailability of certain files linked to flaky tests mentioned in the IDOFT dataset, either due to merged commits or non-existent files in our local system. Consequently, from these 5 successfully executed projects, we collected results of all their flaky tests (35 test cases). We have provided details of these projects in our replication package.
% This is shown in the Table~\ref{tab:execution_resultsRQ3} in Appendix~\ref{app:frequency_of_tests_GPT_dataset}}

\subsection{Experimental Design}
This subsection covers the baselines, code model training and testing, GPT model's prompt template, and execution of the repaired flaky tests.
\subsubsection{Baseline}
Automatically defining different categories of flaky test fixes using only test code and then labeling flaky tests accordingly along with generating a full repair of flaky tests is a task that has not yet been studied before. Previous work has either analyzed the causes of flakiness~\cite{akli2023flakycat} or manually identified the fix for flaky tests~\cite{parry2022developer}. 
To design our experiment, we only focused on code models for prediction, as explained in Section~\ref{Prediction with Code Models}. As our technique relies exclusively on test code, we cannot compare it with traditional machine-learning classifiers, which require a pre-defined set of features such as execution history, coverage information in production code, and static production code features~\cite{fatima2022flakify}. 
%Additionally, given the non-deterministic behavior of flaky test cases, it can be challenging to identify a complete set of features that could potentially be associated with test flakiness and help in suggesting a fix.
Alternatively, only relying on predefined static features from test code is challenging. These features may not capture all the semantics of the code that contribute towards flakiness and may not be good indicators to predict fix categories. Lastly, defining good features, often referred to as feature engineering~\cite{verdonck2021special}, is not an easy task. In contrast, we aim to develop a solution that is practical where testers just pass the test code directly to a model as input and get information regarding their fix category, rather than going through the process of defining features. 

We excluded traditional sequence learning models as baselines, such as the basic RNN or LSTM models, since they are not pre-trained like code models. Hence, they require large datasets to train their many parameters and avoid over-fitting~\cite{pascanu2013difficulty,collobert2011natural}.

Therefore, we evaluated our idea using two SOTA code models, namely  CodeBERT and UniXcoder, both with and without FSL. FSL has been recommended in previous work for small datasets (e.g., for flakiness root-cause prediction~\cite{akli2023flakycat}), but has not been compared with code models without FSL. Thus we also included this experiment.

\subsubsection{Training and Testing Prediction Models}
To evaluate the performance of our code models for predicting the fix categories, we employed a stratified 4-fold cross-validation procedure to ensure unbiased train-test splitting. This procedure involves dividing the dataset into four equal parts, with three parts allocated for training the model and one part held out as the test set. 
%Specifically, we utilized 75\% of the available test cases for training the model and 25\% for testing in each of the four folds. 
Moreover, to prevent over-fitting and under-fitting during the model's training, we employed a validation split, which consists of 30\% of the training data. This validation dataset is used for fine-tuning code models using FNN and a Siamese Network.
We deliberately did not use a higher number of folds since for smaller categories that would not leave enough samples in the train and test sets. For instance, the~\textit{Change Data Format} fix category has only thirty-three positive examples, which would lead to only four positive examples in each fold of a 10-fold cross-validation. 

\subsubsection{Prompt Design}\label{prompt_design}
\begin{figure}[htbp]
    \centering
    % \flakyTestResolutionwithoutLabel
    \begin{tikzpicture}[node distance=0.45cm] % adjust vertical distance
        \node [draw, rectangle, fill=white, text width=7.0cm, align=left, font=\ttfamily, outer sep=0pt, inner sep=5pt] (rect1) {
            \faPencil* \hspace{0.1cm}  \textbf{Input(Prompt):} \\
                         This test case is Flaky: 
            \textbf{[Flaky Code]}.
            Just Provide the fixed code of this test case only to make it non-flaky. Do not provide any other text description.
        };
        \node [below=of rect1, outer sep=0pt, inner sep=0pt] (img) {\includegraphics[width=4cm]{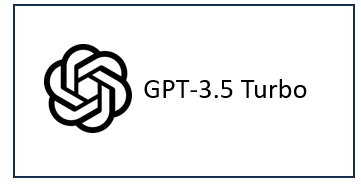}};
        \node [below=of img, draw, rectangle, fill=white, text width=7cm, align=left, font=\ttfamily, outer sep=0pt, inner sep=5pt] (rect2) {
            \faPencil* \hspace{0.1cm} \textbf{Output(Generated Output):} 
            \textbf{[Fixed Flaky Test Code]}
        };
        \draw[->, thick] (rect1.south) -- (img.north);
        \draw[->, thick] (img.south) -- (rect2.north);
    \end{tikzpicture}
   \caption{Flaky test fix using GPT-3.5 Turbo without fix category label.}
    \label{fig:prompting_withoutlabels}
\end{figure}

To get the fixed flaky test code from GPT, we use three different types of prompts, as follows: (1) we pass the flaky test code in the prompt to the model and ask it to generate its fix as shown in Figure~\ref{fig:prompting_withoutlabels}, (2) we provide both the flaky test code and the fix category label we predicted for that flaky test. We then ask the GPT model to generate a fix as shown in Figure~\ref{fig:prompting_labels}, and (3) we extend the content of the second type by also passing a few examples to the model as shown in Figure~\ref{fig:prompting_incontext}. This type of prompting aims to assess GPT's performance with in-context learning. We focused on three fix categories comprising more than 30 instances (to have large enough sample sizes to make meaningful conclusions) within our dataset of 181 tests: \textit{Change Assertion} (64 examples), \textit{Change Data Structure} (41 examples), and \textit{Change Condition} (41 examples). This resulted in a dataset of 146 tests. From these, we randomly selected 5 examples from each category for in-context learning, leaving a total of 131 tests for our in-context learning experiments.

We used these selected examples of flaky tests, along with their respective fixes and fix category labels, as prompts for the GPT model. Our prompt requested the model to provide a fix for a new flaky test based on a given fix category label, utilizing information derived from these example tests and their associated fix category labels.
Our prompt explicitly requests the model to provide only the fixed test code, excluding any additional textual descriptions. This approach streamlines the output, ensuring that the model generates the test code directly without requiring post-processing. In the future, we plan to further test our GPT model using a larger number of examples of flaky tests. This expansion aims to evaluate and enhance the model's capabilities in handling varied scenarios. 
\begin{figure}[htbp]
    \centering
    \flakyTestResolution % Second instance of the TikZ picture
    \caption{Flaky test fix using GPT-3.5 Turbo with fix category label.}
    \label{fig:prompting_labels}
\end{figure}

\begin{figure}[htbp]
    \centering
    \flakyTestResolutionincontextLearning % Second instance of the TikZ picture
    \caption{Flaky test fix using GPT-3.5 Turbo with in-context learning.}
    \label{fig:prompting_incontext}
\end{figure}

\subsubsection{Test Case Execution}\label{ref:testcase_execution}
To verify the automatically generated repairs, as described in Section~\ref{ref:Dataselection_execution}, we ran 35 tests obtained from 5 projects that could be executed. To do so, initially, we executed the original developer-repaired versions of the 35 tests, 10 times each, to help verify the approved fix was indeed correct. Then, we executed the LLM-generated versions of these tests, 10 times each as well. Given that a test may remain flaky even after repair if the correction is inadequate, multiple runs were necessary to ascertain consistent outcomes across executions. That is, a passing or failing test should consistently pass or fail, respectively, in all 10 runs. We have included detailed instructions for executing these tests, along with the corresponding results, within our replication package.

\subsubsection{Evaluation Metrics}
To assess the effectiveness of our classifier, we employed standard evaluation metrics, including precision, recall, and F1-Score, which have been widely used in previous studies of ML-based solutions for flaky tests~\cite{fatima2022flakify,akli2023flakycat,alshammari2021flakeflagger,camara2021vocabulary,pinto2020vocabulary}. 
In our case, precision is of utmost importance, and we emphasize its role in our results. In practical terms, once a fix category is predicted for a flaky test, such as \textit{Change Assertion}, we want testers to confidently proceed with fixing the issue by focusing on the code statements matching the category, e.g., by modifying the assertion type used. Accurately predicting the correct fix category is therefore critical since attempting to fix a test incorrectly can result in wasted time and resources~\cite{micco2016flaky} and may even cause additional failures in the test suite~\cite{habchi2022qualitative,thung2012automatic}.

We used Fisher's Exact test~\cite{raymond1995exact} to assess whether there was a significant difference in performance among the four prediction methods: fine-tuned CodeBERT and UniXcoder with and without FSL, across all nine fix categories. 
Furthermore, in RQ1, we used the Kappa score~\cite{mchugh2012interrater} to evaluate the level of agreement between the researchers who initially labeled the dataset. By using the Kappa score, we were able to assess the reliability and consistency of the initial labels across different categories of fix, despite their varying sample sizes.
Additionally, in RQ3, to evaluate how useful our classification is for improving an automated repair model (i.e., GPT in this case), we have used several metrics from the BLEU score family ~\cite{papineni2002bleu} (i.e., Corpus BLEU, Sentence BLEU, and CodeBLEU), which are widely used metrics in automated code generation and repair studies~\cite{dong2023codescore,abdollahpour2023automatic,liu2023improving,mashhadi2021applying}.

Particularly, CodeBLEU is crucial as it captures both the semantics and syntax of the code, providing a more precise evaluation of the generated code. Unlike the traditional BLEU score, which assesses the similarity between candidate and reference texts, based solely on sequences of \(n\) contiguous tokens (n-grams) overlapping, CodeBLEU incorporates additional components to account for the specific code characteristics. It is defined as a weighted combination of four parts~\cite{ren2020codebleu} as shown in the below equation. 
\begin{equation}
\begin{split}
\text{CodeBLEU} =
\alpha \cdot \text{BLEU} + \beta \cdot \text{BLEU}_{\text{weight}} \\ + \gamma \cdot \text{Match}_{\text{ast}} + \delta \cdot \text{Match}_{\text{df}}
\label{equation1}
\end{split}
\end{equation}
where:
\begin{itemize}
    \item $\text{BLEU}$ represents the standard BLEU score, measuring the overlap of n-grams between the generated and reference code.
   \item $\text{BLEU}_{\text{weight}}$ denotes the weighted n-gram matching, assigning varying weights to n-grams to emphasize important code keywords like \textit{int}, \textit{char}, \textit{public}, and others. Different weights are assigned to different n-grams, with keywords often receiving higher weights.
    \item $\text{Match}_{\text{ast}}$ signifies the syntactic AST match, focusing on the syntactic structure of the generated code.
\item$\text{Match}_{\text{df}}$ assesses the semantic similarity between generated and reference code by examining their data-flow representations. Code is represented as a graph where nodes represent variables, and edges indicate how variable values flow. 
%For example, consider two functions: one returns the value of variable \textit{x}, while the other returns the value of variable \textit{y}. 
Unlike Abstract Syntax Trees (AST), data-flow graphs illustrate how variables interact within the code. We are thus able to measure the semantic match between the candidate and reference code.
\end{itemize}
The parameters  $\alpha$, $\beta$, $\gamma$, and $\delta$ in Equation~\ref{equation1} represent the weights of the above-mentioned four parts in the CodeBLEU score. We have used a value of 0.25 for all four weights, as used by Ren et al.~\cite{ren2020codebleu}. These four components collectively evaluate both the grammatical correctness and logical correctness of the generated code, making CodeBLEU a comprehensive metric for code evaluation.

Beside syntactic similarity metrics like CodeBLEU, we also executed a subset of repaired tests to assess the correctness of GPT-repaired flaky tests. Analysing the execution results, we calculated the passing rate of the repaired tests. Then using bootstrapping~\cite{mooney1993bootstrapping} with a 95\% confidence level, a statistical method for estimating properties of a large sample from repeated sampling (with replacement) from a small sample, we established lower and upper bound percentages of passing test cases within the entire flaky test set, including non-executable test cases. 
Additionally, we used a logistic regression \cite{hosmer2013applied} model to predict the pass rates of non-executable tests based on their CodeBLEU scores in various prompt settings. 

Furthermore, for passing tests, we manually ensured they did not introduce regressions regarding the test class's bug-finding capability. While tests with a CodeBLEU score of 100\% didn't need equivalence checks, for those with scores below that, we manually verified behavioral equivalence with the ground truth. This was possible given the relatively low complexity of the test cases. 

Regarding failing tests, we calculated the Levenshtein Edit Distance~\cite{ristad1998learning} between the generated and actual fixes. This is a measure of similarity between two strings that determines the minimum number of token edits needed to transform one into the other. We can then compute the number and percentage of tokens needing change to achieve perfect alignment with the original developer-repaired test.

Lastly, We employed the Wilcoxon signed-rank~\cite{woolson2007wilcoxon} test to gauge the significance in performance differences of the GPT model across various input prompts.

\subsection{Results}
\subsubsection{RQ1 Results}
To start, 100 test cases were randomly selected and independently labeled by two researchers (that are among the co-authors of this research).The Kappa score was found to be 0.65, indicating a moderate level of agreement between the two raters. Disagreements betwe~en them were eventually resolved with multiple rounds of discussions to finalize the fix categories that are used by the automated labeling tool. 

To evaluate the labeling accuracy of our automated tool, we took another random set of 200 flaky tests and obtained their fix category labels from the tool. The same two researchers then manually labeled these tests, enabling us to calculate the accuracy of our automated labeling tool as shown in Figure~\ref{fig:Labelling_process}. We found that the tool had an accuracy of 98.5\%, with only three tests being mislabeled with multiple fix categories, including the correct one. 

\begin{tcolorbox}[colback=white,colframe=black]
\textbf{RQ1 summary:} 
Our proposed automated labeling system for
flaky tests was evaluated and yielded an accuracy of
98.5\% in correctly labeling a test set of 200 tests
with various fix categories. Having an automated and
accurate labeling tool is essential for this study, future
research, and in practice when building such prediction
models in context.
\end{tcolorbox}
\subsubsection{RQ2 Results}
Table~\ref{tab:Results_RQ2} presents the prediction results for all nine categories of flaky test fixes using four distinct approaches, namely CodeBERT and UniXcoder, fine-tuned with Siamese Networks or FNN. In the former case, we evaluated both language models using Few-Shot Learning (FSL).

UniXcoder without FSL demonstrated superior precision and similar recall in the \textit{Change Assertion}, \textit{Change Data Structure}, \textit{Handle Exception}, \textit{Change Condition}, and \textit{Reorder Parameters} categories, with precision of 96\%, 89\%,  98\%, 87\%, and 91\%, respectively, outperforming the other three approaches. For \textit{Reorder Data}, again UniXcoder without FSL achieves the highest precision score of 96\%. However, when incorporating FSL, UniXcoder demonstrates a superior recall of 88\%. Furthermore, CodeBERT without FSL showed slightly better results for the \textit{Miscellaneous} category, with 83\% precision and 67\% recall, though the difference with UniXcoder is not significant. For the \textit{Reset Variable} category, UniXcoder with FSL yielded high accuracy with 96\% precision and 97\% recall rates, while CodeBERT without FSL achieved a slightly higher recall of 98\% albeit with a low precision of 88\%. Finally, for \textit{Change Data Format}, CodeBERT with FSL outperformed all other approaches with a 97\% precision and recall. However, again the difference with UniXcoder is not significant. 

The \textit{Change Data Format}, \textit{Handle Exception}, \textit{Reset Variable}, \textit{Reorder Data } and \textit{Change Assertion} categories were the easiest for all four approaches to classify, with an average precision score (across all approaches) of 93\%, 96\%, 92\%, 88\%, and 90\%, respectively. This is likely due to the high frequency of common keywords in test code for exception handling, assert statements, resetting variables, and reordering data (with keywords like "sort" and "order by"). In the case of the \textit{Change Data Format} category, the model looks for a long complicated string that is being passed to a function. These strings are easier for the model to identify because we observed that all the flaky tests matching this category contain a long complicated string that needs to be updated to remove flakiness. All four models demonstrate strong performance across most fix categories, exhibiting high recall and precision. However, in the Miscellaneous Category, their performance notably decreases, averaging at 62\%. This lower performance may be attributed to the absence of specific common patterns or code keywords across all tests within this category, which cannot be expected to be homogeneous. Moreover, in the Reorder Parameter Category, CodeBERT, with and without FSL, exhibits significantly lower precision compared to UniXcoder. 
%This lack of consistent elements is challenging for the models in accurately classifying items into the Miscellaneous category. 
Further details about the models' misclassifications are provided in Section~\ref{Discussion}.

Overall, the results presented in Table~\ref{tab:Results_RQ2} suggest that, for most of the fix categories, UniXcoder outperforms CodeBERT significantly, based on a Fisher's Exact test with $\alpha=0.05$.
However, FSL did not significantly improve UniXcoder's performance, which is probably due to the limited dataset with only a small number of positive instances for most fix categories, which hindered fine-tuning. 
\begin{tcolorbox}[colback=white,colframe=black]
\textbf{RQ2 summary:} 
UniXcoder outperformed CodeBERT
for predicting most of the fix categories, with higher
precision and recall. The use of FSL does not significantly improve the prediction accuracy of UniXcoder.
\end{tcolorbox}

\begin{table*}[!htpb]
\setlength{\tabcolsep}{0.3em}
\renewcommand{\arraystretch}{1.5}
\centering
\resizebox{1.0\textwidth}{!}{
\begin{tabular}{|l|lclc|lclc|lclc|}
\hline
Metric   & \multicolumn{4}{c|}{~\textbf{Precision} (\%)}                                                                                              & \multicolumn{4}{c|}{~\textbf{Recall} (\%)}                                                                                                          & \multicolumn{4}{c|}{~\textbf{F1} (\%)}                                                                                                              \\ \hline
Category & \multicolumn{1}{l|}{CB} & \multicolumn{1}{l|}{CB with FSL} & \multicolumn{1}{l|}{UC}          & \multicolumn{1}{l|}{UC with FSL} & \multicolumn{1}{l|}{CB}          & \multicolumn{1}{l|}{CB with FSL} & \multicolumn{1}{l|}{UC}          & \multicolumn{1}{l|}{UC with FSL} & \multicolumn{1}{l|}{CB}          & \multicolumn{1}{l|}{CB with FSL} & \multicolumn{1}{l|}{UC}          & \multicolumn{1}{l|}{UC with FSL} \\ \hline
CA       & \multicolumn{1}{c|}{88} & \multicolumn{1}{c|}{86}          & \multicolumn{1}{c|}{\textbf{96}} & 93                               & \multicolumn{1}{l|}{88}          & \multicolumn{1}{c|}{\textbf{86}} & \multicolumn{1}{l|}{\textbf{86}} & 85                               & \multicolumn{1}{l|}{88}          & \multicolumn{1}{c|}{86}          & \multicolumn{1}{l|}{\textbf{91}} & 89                               \\ \hline
CDS      & \multicolumn{1}{c|}{86} & \multicolumn{1}{c|}{76}          & \multicolumn{1}{c|}{\textbf{89}} & 85                               & \multicolumn{1}{l|}{86}          & \multicolumn{1}{c|}{71}          & \multicolumn{1}{l|}{\textbf{87}} & 81                               & \multicolumn{1}{l|}{87}          & \multicolumn{1}{c|}{74}          & \multicolumn{1}{l|}{\textbf{88}} & 83                               \\ \hline
HE       & \multicolumn{1}{l|}{97} & \multicolumn{1}{c|}{95}          & \multicolumn{1}{l|}{\textbf{98}} & 95                               & \multicolumn{1}{l|}{87}          & \multicolumn{1}{c|}{88}          & \multicolumn{1}{l|}{89}          & \textbf{90}                      & \multicolumn{1}{l|}{92}          & \multicolumn{1}{c|}{91}          & \multicolumn{1}{l|}{\textbf{93}} & \textbf{93}                      \\ \hline
RV       & \multicolumn{1}{l|}{88} & \multicolumn{1}{c|}{91}          & \multicolumn{1}{l|}{94}          & \textbf{96}                      & \multicolumn{1}{l|}{\textbf{98}} & \multicolumn{1}{c|}{95}          & \multicolumn{1}{l|}{97}          & 96                               & \multicolumn{1}{l|}{93}          & \multicolumn{1}{c|}{93}          & \multicolumn{1}{l|}{\textbf{96}} & \textbf{96}                      \\ \hline
CC       & \multicolumn{1}{l|}{91} & \multicolumn{1}{c|}{84}          & \multicolumn{1}{l|}{\textbf{87}} & 74                               & \multicolumn{1}{l|}{84}          & \multicolumn{1}{c|}{70}          & \multicolumn{1}{l|}{\textbf{87}} & 70                               & \multicolumn{1}{l|}{\textbf{87}} & \multicolumn{1}{c|}{76}          & \multicolumn{1}{l|}{\textbf{87}} & 72                               \\ \hline
RD       & \multicolumn{1}{l|}{88} & \multicolumn{1}{c|}{80}          & \multicolumn{1}{l|}{\textbf{96}} & 90                               & \multicolumn{1}{l|}{82}          & \multicolumn{1}{c|}{\textbf{88}} & \multicolumn{1}{l|}{72}          & \textbf{88}                      & \multicolumn{1}{l|}{85}          & \multicolumn{1}{c|}{84}          & \multicolumn{1}{l|}{82}          & \textbf{89}                      \\ \hline
CDF      & \multicolumn{1}{l|}{96} & \multicolumn{1}{c|}{\textbf{97}} & \multicolumn{1}{l|}{90}          & 90                               & \multicolumn{1}{l|}{93}          & \multicolumn{1}{c|}{\textbf{97}} & \multicolumn{1}{l|}{94}          & 90                               & \multicolumn{1}{l|}{96}          & \multicolumn{1}{c|}{\textbf{97}} & \multicolumn{1}{l|}{93}          & 95                               \\ \hline
MC       & \multicolumn{1}{l|}{\textbf{83}} & \multicolumn{1}{c|}{79}          & \multicolumn{1}{l|}{82}          & 76                               & \multicolumn{1}{l|}{\textbf{67}}        & \multicolumn{1}{c|}{61}          & \multicolumn{1}{l|}{62}          & 61                               & \multicolumn{1}{l|}{\textbf{74}} & \multicolumn{1}{c|}{\textbf{70}} & \multicolumn{1}{l|}{70}          & 68                               \\ \hline
RP       & \multicolumn{1}{l|}{53} & \multicolumn{1}{c|}{48} & \multicolumn{1}{l|}{\textbf{91}}          & 88                               & \multicolumn{1}{l|}{\textbf{94}}          & \multicolumn{1}{c|}{88}          & \multicolumn{1}{l|}{\textbf{91}} & 82                               & \multicolumn{1}{l|}{64}          & \multicolumn{1}{c|}{68}          & \multicolumn{1}{l|}{\textbf{91}} & 85                               \\ \hline
\end{tabular}

    }
    \caption{Comparison of the prediction results for each fix category using CodeBERT (CB) and UniXcoder (UC), both with and without FSL. The highest value per metric for each fix category is highlighted.}
    % UniXcoder with and without FSL outperforms CodeBERT with higher precision for Change Assertion (CA), Change Data Structure (CDS),  Reset Variable (RV), Change Condition (CC), Reorder Data (RD) and  Reorder Parameters (RP)  fix categories. CodeBERT without FSL and with FSL shows a slightly better F1 score than UnixCoder for the Miscellaneous (MC) and Change Data Format (CDF) category (CDF), respectively.   }
\label{tab:Results_RQ2}
% \end{table*}
\end{table*}

\subsubsection{RQ3 Results}~\label{RQ3_Results}
Table~\ref{tab:Results_RQ3} reports on the evaluation results generated by GPT 3.5 Turbo, which processes a flaky test and provides its corrected version, with and without our predicted fix category labels as input. The assessment metrics include median and mean scores for Corpus BLEU, Sentence BLEU, and CodeBLEU across the 181 flaky tests in the test set.
The inclusion of our predicted fix category labels, in the prompt, increases the median of all three metrics by 7\%, 6\%, and 4\% for Corpus, Sentence, and CodeBLEU, respectively. The improvements are statistically significant based on non-parametric Wilcoxon signed-rank tests at a significance level of $\alpha=0.05$, with a p-value below 0.0001 in all three cases. %This p-value falls below the specified threshold for all three evaluation metrics: CodeBLEU score, Sentence BLEU, and Corpus BLEU.
This improvement is attributed to the additional information provided to the GPT model, aiding it in precisely identifying and addressing the issues causing test flakiness.
This finding also indirectly suggests that our predicted fix categories are well defined and may effectively help testers identify the specific code segments that require modification to mitigate flakiness. 
In the rest of this RQ, we only report CodeBLEU scores, since they are designed to more directly assess the semantics and syntax of code rather than natural language text, thus providing a more accurate evaluation of generated test code.

Table~\ref{tab:mean_codeBLEUSCore} shows median and mean CodeBLEU scores per fix category, with and without using label information.
As we can see, in most categories, the labels improve the mean and median CodeBLEU scores. The improvements per category range from 0\% to 17\% for the mean and from -1\% to 22\% for the median CodeBLEU. Smaller improvements are either in categories with very few samples (e.g., Reset Variables with only one sample) or when the results are already quite high (e.g., Change Condition with a median CodeBLEU of 89\% without labels). The only category where the Median results slightly dropped (1\%) after providing the labels is the Misc category. This is most likely because the \textit{Misc} label does not offer practical information on how to resolve the flakiness.

Now to expand the analysis to the third prompting approach (in-context learning), Table~\ref{tab:Incontextlearning_results} compares the three approaches for the \textit{Change Assertion}, \textit{Change Data Structure}, and \textit{Change Condition} fix categories. As explained, we only look at these categories since after removing 5 samples per category to be included in the context, only these categories have enough samples to do a proper analysis  and run statistical tests.  

\begin{figure*}[!htb]
    \centering
    \subfloat[\centering Change Assertion]{{\includegraphics[width=0.66\linewidth]{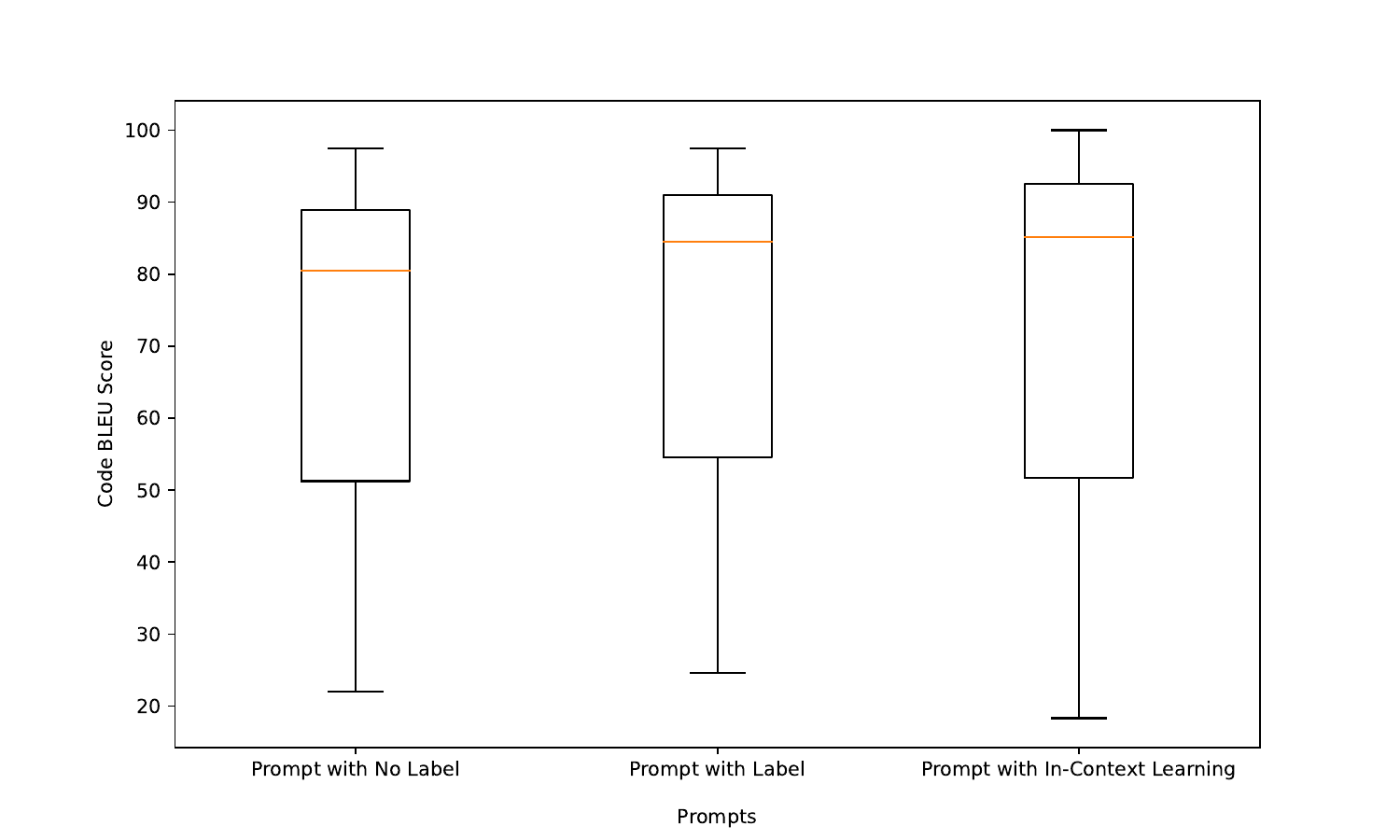} }\label{changeassertion_dist}}%
    \hfill
    \subfloat[\centering Change Data Structure]{{\includegraphics[width=0.66\linewidth]{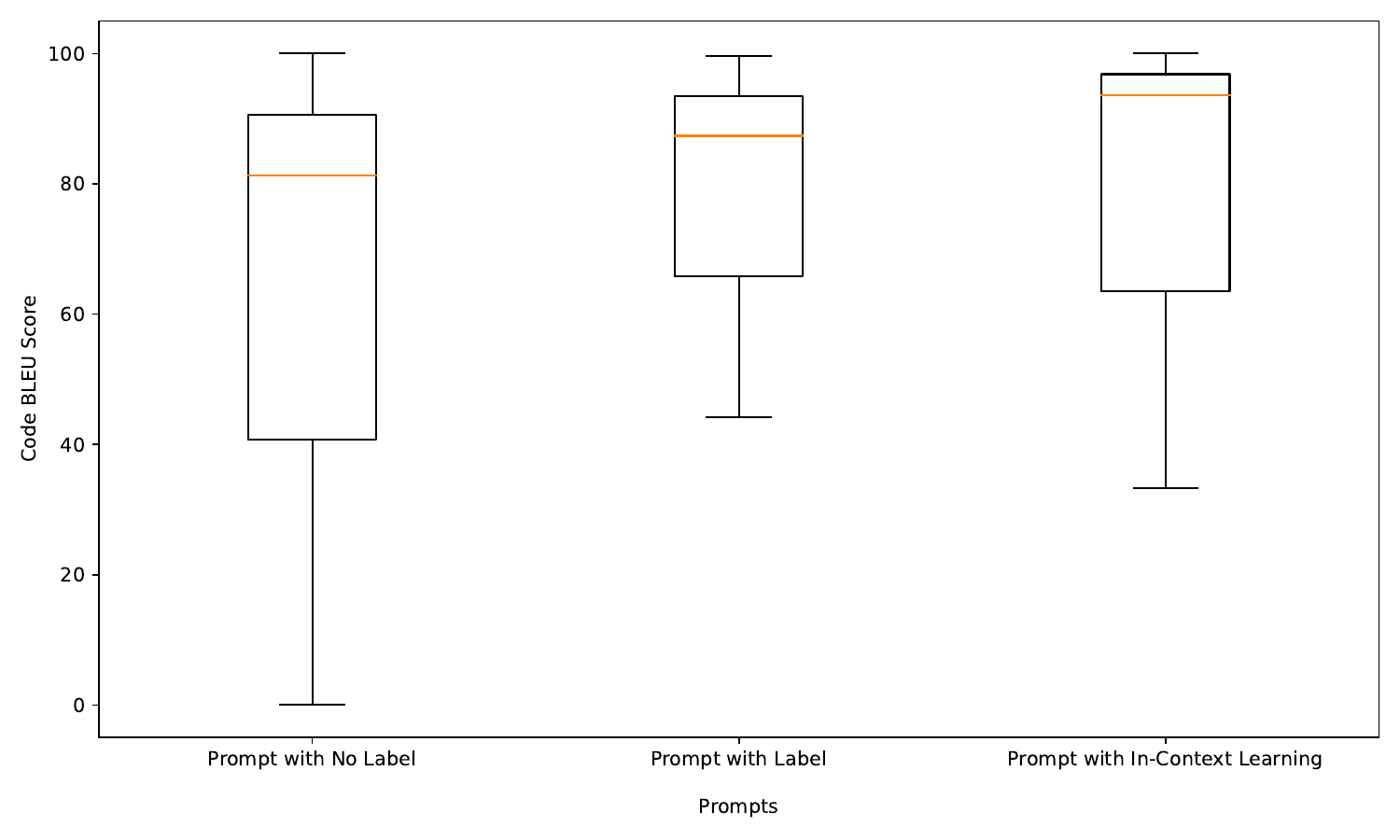} }\label{changedatastructuredist}}%
    \hfill
    \subfloat[\centering Change Condition]{{\includegraphics[width=0.66\linewidth]{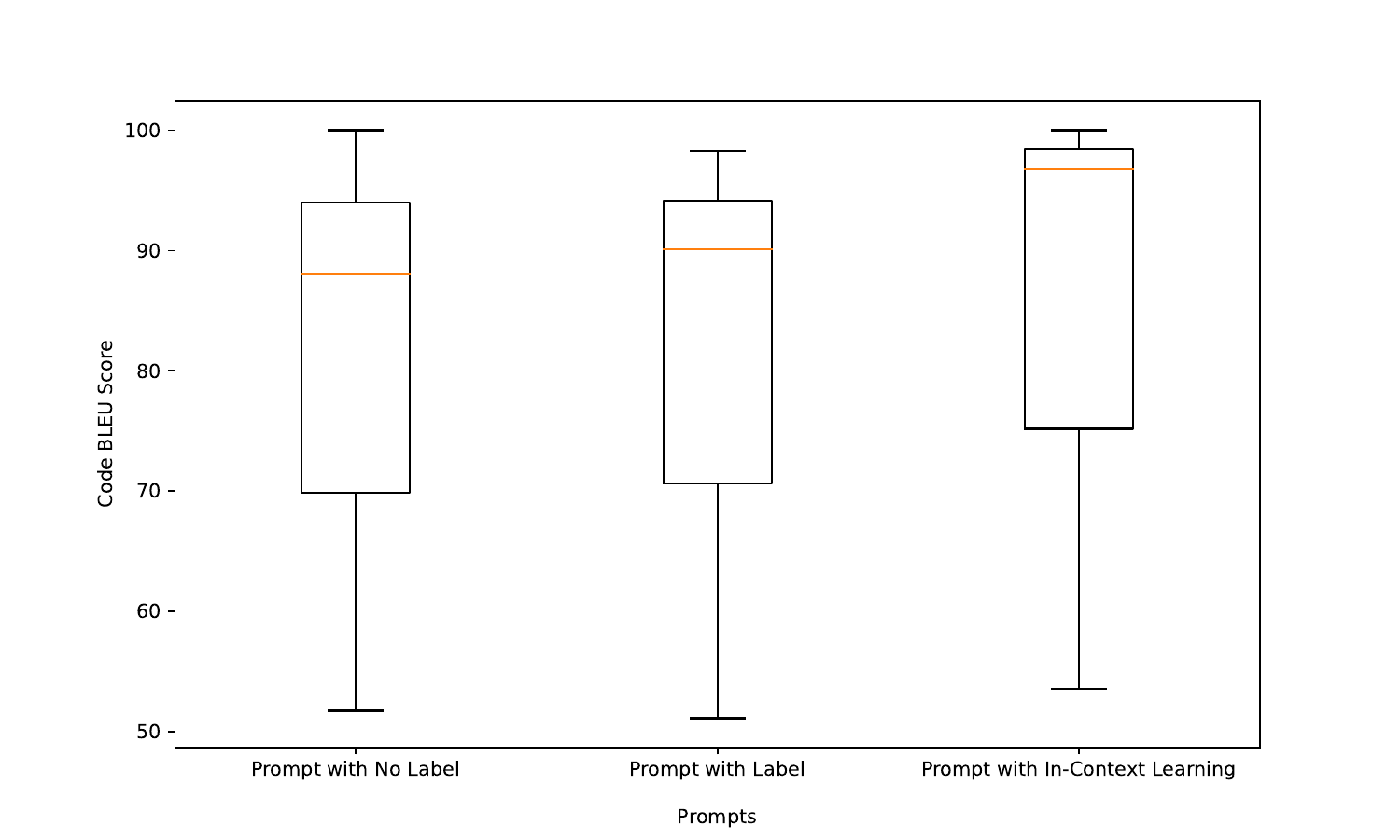} }\label{changecondition_dist}}%
     
    \caption{Distribution of CodeBLEU scores with respect to three prompts.}%
    \label{fig:DistributionofCodeBleu}%
\end{figure*}

Looking at the table, we see that, for the~\textit{Change Data Structure} category, the mean and median CodeBLEU scores increase from 80\% to 86\% and 87\% to 94\%, respectively, when we use in-context learning versus simply providing the labels.
Similarly, for the ~\textit{Change Condition} category, mean and median CodeBLEU scores increase from 88\% to 92\% and 90\% to 97\%, respectively.
However, in the case of the~\textit{Change Assertion} category, there is no discernible improvement in the model's performance after exposure to a limited number of examples. 
This could be attributed to the variability of assertion statements within the test code, suggesting that multiple approaches may exist to fix assertions and resolve flakiness. 
It is plausible that the number and variety of examples provided was insufficient for the model to grasp the diverse spectrum of assertion-related fixes. 
Further, as depicted in the boxplots in Figure ~\ref{fig:DistributionofCodeBleu}, the overall distribution of the CodeBLEU scores shows significant variance for all types of prompts and the three fix categories. 
The low minimum values in all cases shows that there are flaky tests that are not fixed regardless of the prompting strategy.
However, when looking at the CodeBLEU over all three categories for in-context learning, it is clear that, for the majority of the cases, the CodeBLEU scores are quite high. For example, even for the Change Assertion category, which has the lowest minimum scores, 75\% of samples have CodeBLEUs higher than 55\%, and for half of the test cases the CodeBLEU scores are greater than 85\%.

Given the variance in CodeBLEU values per category, we again used the non-parametric Wilcoxon signed-rank test with a significance level of $\alpha=0.05$ to make sure the improvements we see in medians are statistically significant as well. Our findings indicate that, when accounting for all three categories, both for prompts with label and without label,~in-context learning makes a significant difference, with p-values below the significance threshold \( p<0.001 \).
However, looking at individual categories, this difference for prompts with label is only significant for \emph{Change Data Structure} and \emph{Change Condition}, with p-values \( p<0.001 \) and \( p=0.007 \), respectively.

% both for prompts with label and without label, in-context learning makes a significant difference, with p-values below the significance threshold $(P < .001$).
% However, looking at individual categories, this difference for prompts with label is only significant for ~\emph{Change Data Structure} and ~\emph{Change Condition}, with p-values $(P=.00024)$ and $(P=.0072)$, respectively.
In future investigations, we intend to systematically explore the use of different prompts by augmenting the number and types of examples or experimenting with other prompt engineering strategies~\cite{liu2023improving}.
Further details regarding GPT's performance on individual fix categories are discussed in Section \ref{Discussion}.

%Tests Execution results
To gain deeper insights into the performance of GPT-repaired flaky tests, we conducted a series of evaluations based on test execution results. For the reasons explained in Section~\ref{ref:Dataselection_execution}, we could only execute a total of 35 tests, each repeated 10 times, resulting in 24 passing and 11 failing tests, with consistent results across the 10 runs. 

From the 35 executable tests, generated with labels and examples, for the three categories where the latter are available, we would like to first estimate the passing rate of fixed tests while accounting for the uncertainty associated with such a limited sample. Using Bootstrapping, we estimate a 95\% confidence interval for the percentage of passing tests, that ranges from 51\% to 83\%. This interval is informative for testers, who do not have access to CodeBLEU scores in practical testing scenarios, and would like to know what to expect from \ourapproach. Based on these results, we conclude that \ourapproach ~provides an effective solution in a large proportion of cases. 

\begin{figure*}[!htb]
    \centering
    \includegraphics[width=1.0\linewidth]{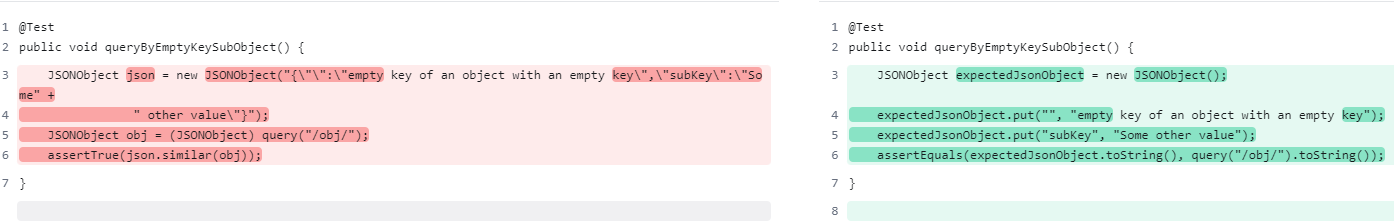}
    \caption{Comparison between a GPT-repaired flaky test (right) and its original developer repair (left), demonstrating a case where the GPT-repaired test passed upon execution despite displaying a very low CodeBLEU score of 36\%.}
    \label{passed_test_lowCodeBLEU}
\end{figure*}

% The main analysis we are interested in is to find 
Next we are interested in determining the relationship between execution outcomes and the corresponding CodeBLEU scores to ascertain the reliability of the CodeBLEU metric in assessing the generated test. Specifically, we expect that tests with higher CodeBLEU scores to have a higher likelihood of passing and vice-versa.
Overall, the average score among passing and failing tests is 94\% and 74\%, respectively, suggesting that higher CodeBLEU scores are an indicator of passing tests.
In particular, among the passing tests, 16 out of 24 achieved a perfect CodeBLEU score of 100\%, indicating an exact match with the developer-repaired versions. Additionally, 22 tests scored above or equal to 80\% on CodeBLEU. Conversely, among the 11 failed tests, only five tests achieved CodeBLEU scores above or equal to 80\%. However, one passing test exhibited a notably lower CodeBLEU score of 36\%, as illustrated in Figure~\ref{passed_test_lowCodeBLEU}, showing an alternative but correct repair. This particular case perfectly illustrates the limitations of using CodeBLEU scores.  

\begin{figure}[!htb]
\includegraphics[width=1.0\linewidth]{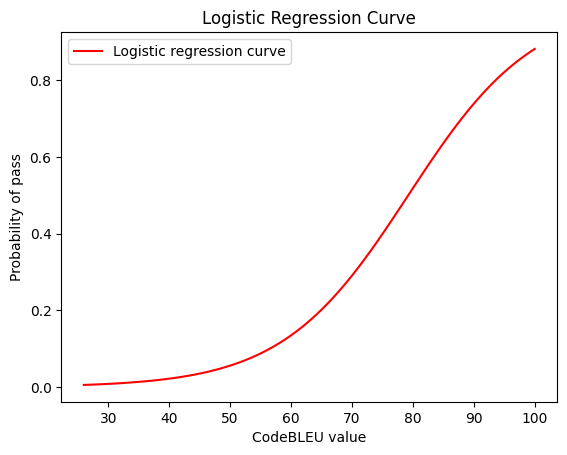}
    \caption{The Logistic Regression Curve showing a relation between CodeBLEU scores and probability of passing.}
    \label{LR_Curve}
\end{figure}

To quantitatively model the relationship between CodeBLEU scores and the test passing probabilities, we train a Logistic Regression (LR)~\cite{hosmer2013applied} model on the 35 executable tests, using their pass/fail labels, achieving an accuracy of 80\% (correctly predicted test cases). The choice of LR is due to the small sample and the fact that we have one explanatory variable, CodeBLEU, that is expected to have a monotonic relationship with the test passing probability. The model's goodness of fit, measured by Chi-Square test~\cite{mchugh2013chi}, yields a p-value of 0.010, which indicates a statistically significant relationship between CodeBLEU scores and the probability for a test to pass, despite the small size of the sample. The fitted LR curve between CodeBLEU scores and the probability for a test to pass is shown in Figure~\ref{LR_Curve}, showing a sharp increase in passing probability beyond CodeBLEU scores of 50.

To estimate the passing test rate among the non-executable tests, we apply this trained LR model to predict the outcomes of these tests. 
We calculate 95\% confidence intervals for the predicted probabilities using standard errors derived via the delta method~\cite{assmann1996confidence}. This method allows us to approximate the variance of the predicted probabilities by first calculating the standard errors on the logit scale and then transforming these to the probability scale. With this, we ensure that the confidence intervals account for the non-linear relationship between the logit and the predicted probability, giving a reliable measure of uncertainty around our predictions.
% We calculate 95\% confidence intervals for the predicted probabilities, using standard errors.
According to our estimates, from the 181 test dataset, we obtain the following probability confidence intervals and passing test estimates:
\begin{itemize}
    \item With No Label prompt:
     Confidence Interval: 0.44–0.50 | 80 (44\%) tests to  91 (50\%) tests are projected to pass. 
    \item With Label prompt: 
    Confidence Interval: 0.51–0.57 |  92 (51\%) tests to 103 (57\%) tests are projected to pass. 

\end{itemize}
From the 131-test dataset (those that we can apply in-context learning on), we obtain the following results:
\begin{itemize}
    \item With No Label prompt: 
    Confidence Interval: 0.44–0.52 | 58 (44\%) tests to 68 (52\%) tests are projected to pass. 

    \item With Label prompt: Confidence Interval: 0.50–0.58 | 66 (50\%) tests to 76 (58\%)  tests are projected to pass. 

    \item With in-context learning: Confidence Interval: 0.53–0.60 | 69 (53\%) tests to 79 (60\%) tests are projected to pass. 
\end{itemize}

Tables~\ref{tab:passing_estimate_181} and~\ref{tab:passing_estimate_131} provide the passing test estimates in terms of upper and lower bounds across different prompt settings, for the 181 and 131 test datasets, respectively.

\begin{figure}[!htb]
    \includegraphics[width=1.02\linewidth]{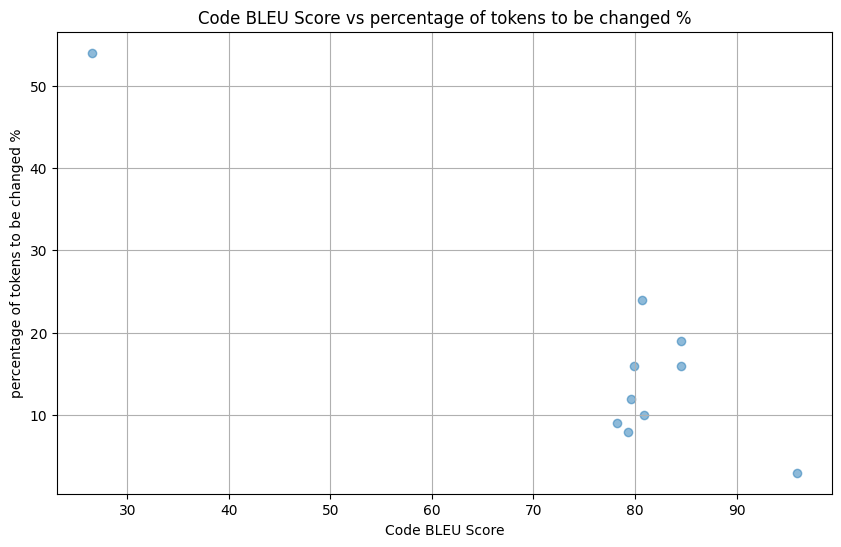}
    \caption{Comparison of CodeBLEU Score with the percentage of tokens changed in failed GPT-generated tests}
    \label{token_percentage_editDistance}
\end{figure}

From the above results, we can conclude that providing labels with the prompt to GPT does indeed make a difference in terms of passing tests. Overall, a reasonable proportion of generated tests with labels are predicted to pass, ranging approximately from 50\% to 60\% when accounting for uncertainty. 
Furthermore, for the 11 GPT-generated tests that failed upon execution, we calculated the edit distance as a surrogate measure to assess the manual effort required for fixing. The results revealed that, on average, 16\% of the tokens needed to be replaced to convert a failed test into the corresponding ground truth.Figure~\ref{token_percentage_editDistance} depicts the trend between CodeBLEU scores and the percentage of token replacement. The scatter plot suggests that GPT-repaired tests with higher CodeBLEU scores tend to require fewer edits for repair, suggesting closer alignment with the developer-repaired versions.
Lastly, to gain further insights into GPT's performance, we analyzed the fix categories of the 11 failed tests. Among these, six tests fell under the \emph{Change Data Format} category. Of the remaining five, three were categorized under \emph{Change Assertion} and two under \emph{Change Data Structure}. The predominance of failed tests in the \emph{Change Data Format} category can be attributed to the absence of in-context learning for this category, due to insufficient available instances as explained in Section~\ref{prompt_design}. Our approach for this category relied solely on prompts with label information, affecting the quality of generations.~However, as visible in Table~\ref{tab:Incontextlearning_results}, for the other three fix categories, the results indicate notable improvements in GPT's generation quality with the implementation of in-context learning.
As mentioned earlier, all generated tests, along with their CodeBLEU scores and execution results, are provided in our replication package.

\begin{tcolorbox}[colback=white,colframe=black]
\textbf{RQ3 summary:} 
In all fix categories, either providing category labels or applying in-context learning significantly boosts the GPT model's ability to repair flaky tests. For two categories, where in-context learning is possible, it provides better results than prompts with labels.
~Furthermore, when providing labels in prompts and with in-context learning, when the latter is possible, we project that a large percentage $(>50\%)$ of repaired tests will pass. For the tests that fail, required changes are rather limited, with an average of 16\% of the tokens to be changed in test code.
\end{tcolorbox}
\begin{table*}[!htpb]
    \setlength{\tabcolsep}{0.6em}
    \renewcommand{\arraystretch}{1.95}
    \centering
    \resizebox{0.60\textwidth}{!}{
        \begin{tabular}{|c|cc|cc|}
            \hline
            \multirow{2}{*}{\textbf{Metrics}} & \multicolumn{2}{c|}{\textbf{Prompts With Labels}} & \multicolumn{2}{c|}{\textbf{Prompts Without Labels}} \\ \cline{2-5} 
            & \multicolumn{1}{c|}{\textbf{Mean}} & \multicolumn{1}{c|}{\textbf{Median}} & \multicolumn{1}{c|}{\textbf{Mean}} & \textbf{Median} \\ \hline
            Corpus BLEU & \multicolumn{1}{c|}{76} & 82 & \multicolumn{1}{c|}{69} & 75 \\ \hline
            Sentence BLEU & \multicolumn{1}{c|}{77} & 82 & \multicolumn{1}{c|}{70} & 76 \\ \hline
            CodeBLEU & \multicolumn{1}{c|}{81} & 85 & \multicolumn{1}{c|}{75} & 81 \\ \hline
        \end{tabular}
    }
    \vspace{0.2em} % Adjust the space between the table and caption as needed
    \caption{Evaluation of the GPT-3.5 generated flaky tests using BLEU Scores across all categories for 181 flaky tests.}
    \label{tab:Results_RQ3}
\end{table*}

\begin{table*}[!htpb]
    \setlength{\tabcolsep}{0.8em}
    \renewcommand{\arraystretch}{1.2}
    \centering
    \resizebox{0.76\textwidth}{!}{
     \fontsize{11}{13.2}\selectfont % Set font size to 11pt
        \begin{tabular}{|c|c|c|c|c|c|}
            \hline
            \multirow{3}{*}{\textbf{\begin{tabular}[c]{@{}c@{}}Fix \\ Categories\end{tabular}}} & \multicolumn{4}{c|}{\textbf{CodeBLEU Score}}                                                                                                                        & \multirow{3}{*}{\textbf{Count}} \\ \cline{2-5}
            & \multicolumn{2}{c|}{\textbf{Prompts With Labels}}                                         & \multicolumn{2}{c|}{\textbf{Prompts Without Labels}}                                     &                                 \\ \cline{2-5}
            & \textbf{Mean} & \textbf{Median} & \textbf{Mean} & \textbf{Median} &                                 \\ \hline
            Change Assertion                                                                    & 80                    & 85              & 76                   & 79                                   & 64                              \\ \hline
            Change Condition                                                                    & 87                    & 90              & 84                   & 89                                   & 41                              \\ \hline
            Change Data Structure                                                               & 87                    & 81              & 72                   & 82                                   & 41                              \\ \hline
            Misc                                                                                & 76                    & 83              & 75                   & 82                                   & 22                              \\ \hline
            Change Data Format                                                                  & 77                    & 80              & 60                   & 58                                   & 13                              \\ \hline
            Reorder Parameters                                                                  & 86                    & 86              & 77                   & 83                                   & 10                              \\ \hline
            Reorder Data                                                                        & 71                    & 73              & 69                   & 73                                   & 5                               \\ \hline
            Handle Exceptions                                                                   & 66                    & 73              & 59                   & 60                                   & 7                               \\ \hline
            Reset Variable                                                                      & 95                    & 95              & 95                   & 95                                   & 1                               \\ \hline
        \end{tabular}
    }
    \vspace{0.2em}  % Adjust the space between table and caption as needed
    \caption{Median and mean CodeBLEU score per fix category, with and without providing the fix category labels in the GPT prompt.}
    \label{tab:mean_codeBLEUSCore}
\end{table*}
\begin{table*}[!htpb]
    \setlength{\tabcolsep}{0.7em}
    \renewcommand{\arraystretch}{1.65}
    \centering
    \resizebox{0.85\textwidth}{!}{
\begin{tabular}{|c|ccccccc|}
\hline
\multirow{3}{*}{\textbf{\begin{tabular}[c]{@{}c@{}}Fix \\ Categories\end{tabular}}} & \multicolumn{7}{c|}{\textbf{CodeBLEU Score}}                                                                                                                                                                                                                                                                                              \\ \cline{2-8} 
                                                                                    & \multicolumn{2}{c|}{\textbf{\begin{tabular}[c]{@{}c@{}}Prompts \\ Without Label\end{tabular}}} & \multicolumn{2}{c|}{\textbf{\begin{tabular}[c]{@{}c@{}}Prompts \\ With Label\end{tabular}}} & \multicolumn{2}{c|}{\textbf{\begin{tabular}[c]{@{}c@{}}Prompts with \\ In-Context Learning\end{tabular}}} & \multirow{2}{*}{\textbf{Count}} \\ \cline{2-7}
                                                                                    & \multicolumn{1}{c|}{\textbf{Mean}}          & \multicolumn{1}{l|}{\textbf{Median}}          & \multicolumn{1}{l|}{\textbf{Mean}}         & \multicolumn{1}{l|}{\textbf{Median}}        & \multicolumn{1}{c|}{\textbf{Mean}}                & \multicolumn{1}{c|}{\textbf{Median}}               &                                 \\ \hline
Change Assertion                                                                    & \multicolumn{1}{c|}{77}                        & \multicolumn{1}{c|}{80}                       & \multicolumn{1}{c|}{79}                       & \multicolumn{1}{c|}{84}                     & \multicolumn{1}{c|}{79}                              & \multicolumn{1}{c|}{85}                            & 59                              \\ \hline
Change Data Structure                                                               & \multicolumn{1}{c|}{75}                        & \multicolumn{1}{c|}{81}                       & \multicolumn{1}{c|}{80}                       & \multicolumn{1}{c|}{87}                     & \multicolumn{1}{c|}{86}                              & \multicolumn{1}{c|}{94}                            & 36                              \\ \hline
Change Condition                                                                    & \multicolumn{1}{c|}{85}                        & \multicolumn{1}{c|}{88}                       & \multicolumn{1}{c|}{88}                       & \multicolumn{1}{c|}{90}                     & \multicolumn{1}{c|}{92}                              & \multicolumn{1}{c|}{97}                            & 36                              \\ \hline
\end{tabular}
    }
    \vspace{0.2em}  % Adjust the space between table and caption as needed
    \caption{Comparison of median and mean CodeBLEU scores for different fix categories using fix category labels and in-context learning.}
    \label{tab:Incontextlearning_results}
\end{table*}

\begin{table*}[!htpb]
    \setlength{\tabcolsep}{0.7em}
    \renewcommand{\arraystretch}{1.65}
    \centering
    \resizebox{0.75\textwidth}{!}{
\begin{tabular}{|c|cc|cc|}
\hline
\multirow{2}{*}{\textbf{Prompts}} & \multicolumn{2}{c|}{\textbf{Lower Bound}} & \multicolumn{2}{c|}{\textbf{Upper Bound}} \\ \cline{2-5}
                                  & \multicolumn{1}{c|}{\textbf{Passing Tests \#}} & \multicolumn{1}{l|}{\textbf{Passing \%}} & \multicolumn{1}{l|}{\textbf{Passing Tests \#}} & \multicolumn{1}{l|}{\textbf{Passing \%}} \\ \hline
With No Label                     & \multicolumn{1}{c|}{80}                        & 44\%                                     & \multicolumn{1}{c|}{91}                        & 50\%                                     \\ \hline
With Label                        & \multicolumn{1}{c|}{92}                        & 51\%                                     & \multicolumn{1}{c|}{103}                       & 57\%                                     \\ \hline
\end{tabular}}
    \vspace{0.2em} % Adjust the space between the table and caption as needed
    \caption{Passing Estimates from the 181 test dataset for the GPT-repaired tests with and without providing the fix category labels in the GPT prompt.}
    \label{tab:passing_estimate_181}
\end{table*}

\begin{table*}[!htpb]
    \setlength{\tabcolsep}{0.7em}
    \renewcommand{\arraystretch}{1.65}
    \centering
    \resizebox{0.75\textwidth}{!}{
    \begin{tabular}{|c|cc|cc|}
        \hline
        \textbf{Prompts} & \multicolumn{2}{c|}{\textbf{Lower Bound}} & \multicolumn{2}{c|}{\textbf{Upper Bound}} \\ \cline{2-5}
                          &  \multicolumn{1}{c|}{\textbf{Passing Tests \#}} & \multicolumn{1}{c|}{\textbf{Passing \%}} &  \multicolumn{1}{c|}{\textbf{Passing Tests \#}} & \multicolumn{1}{c|}{\textbf{Passing \%}} \\ \hline
        With No Label     &  \multicolumn{1}{c|}{58}                    & \multicolumn{1}{c|}{44\%}                &  \multicolumn{1}{c|}{68}                      & \multicolumn{1}{c|}{52\%}                 \\ \hline
        With Label        &  \multicolumn{1}{c|}{66}                    & \multicolumn{1}{c|}{50\%}                 &  \multicolumn{1}{c|}{76}                        & \multicolumn{1}{c|}{58\%}                 \\ \hline
        In-Context Learning &  \multicolumn{1}{c|}{69}                      & \multicolumn{1}{c|}{53\%}                 &  \multicolumn{1}{c|}{79}                        & \multicolumn{1}{c|}{60\%}                 \\ \hline
    \end{tabular}
    }
    \vspace{0.2em} % Adjust the space between the table and caption as needed
    \caption{Passing Estimates for the GPT-repaired tests using fix category labels and in-context learning (for the 131 tests). }
    \label{tab:passing_estimate_131}
\end{table*}


\section{Discussion}\label{Discussion}
In this section, with qualitative analysis, we aim to ascertain not only theoretical soundness but also the practical utility of our solution in addressing the repair of flaky tests. 
%Change here, in reflection of RQ3 results. Also add here the results of wrong generation. add one example. 
\subsection{Qualitative Analysis of the Model's Performance}
To gain a deeper understanding of the best model's performance, we have conducted a qualitative analysis of representative flaky tests from various fix categories. 
\subsubsection{Code Model's Prediction Results for Flaky Test Fix Category }
In this section, we are analyzing sample test cases and their predictions using one approach, namely UniXcoder without FSL, since on mean it performs the best among the prediction models.
We specifically target True Positives (TP) and a False Negatives (FN) from the same category, to provide the reader with insights regarding how these models can predict flaky fix categories. 
We also investigate False Positives (FP), since FP instances waste testers' time when they follow wrong suggestions and therefore directly affects the applicability of the proposed approach.  

\begin{figure}%
    \centering
    {{\includegraphics[width=7.5cm]{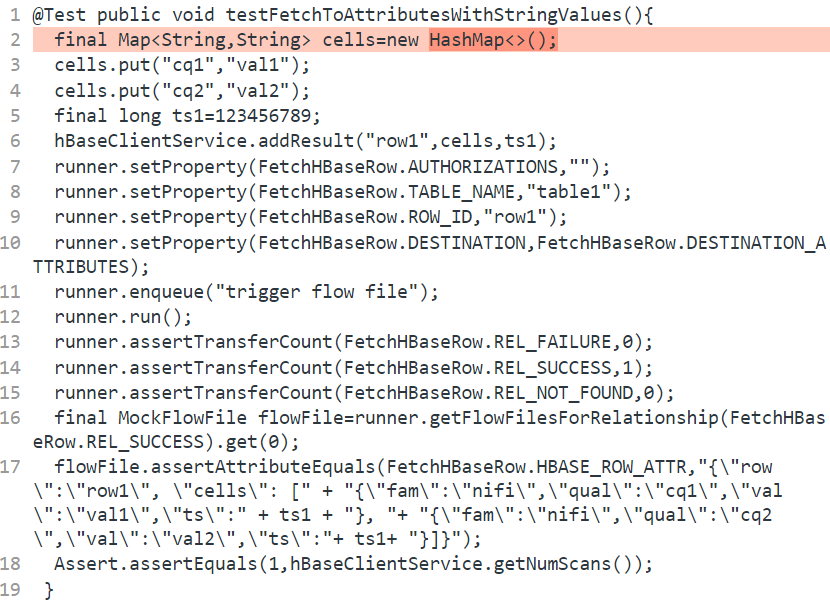} }}%
    \qquad
    % \subfloat[\centering Fixed Flaky Test]{{\includegraphics[width=7cm]{Images/Change_DataStructure_repair_code1} }}%
    \caption{Flaky test example with correct mapping to Change Data Structure fix category as HashMap here should be replaced with LinkedHashMap to remove flakiness.}%
    \label{fig:changedatastructure1}%
\end{figure}

\begin{figure}%
    \centering
    \subfloat[\centering Flaky Test]{{\includegraphics[width=9cm]{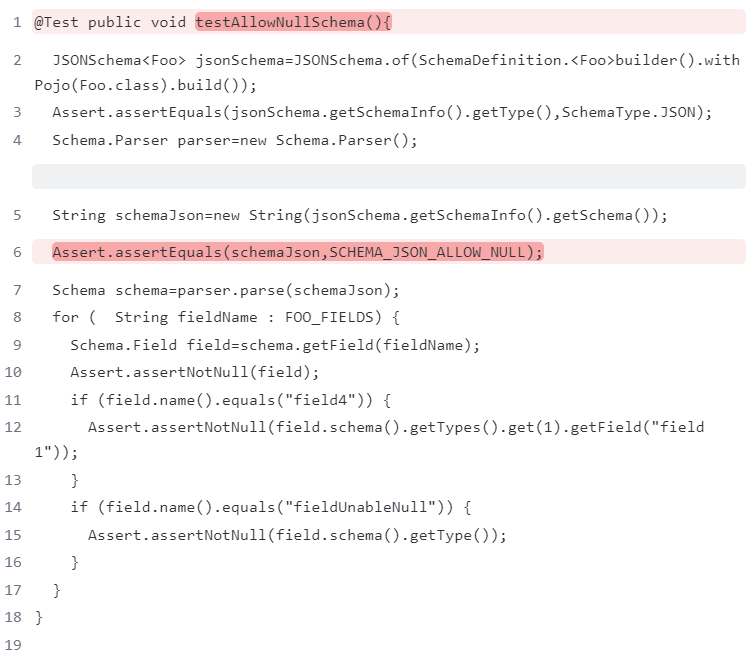} }}%
    \qquad
    \subfloat[\centering Fixed Flaky Test]{{\includegraphics[width=9cm]{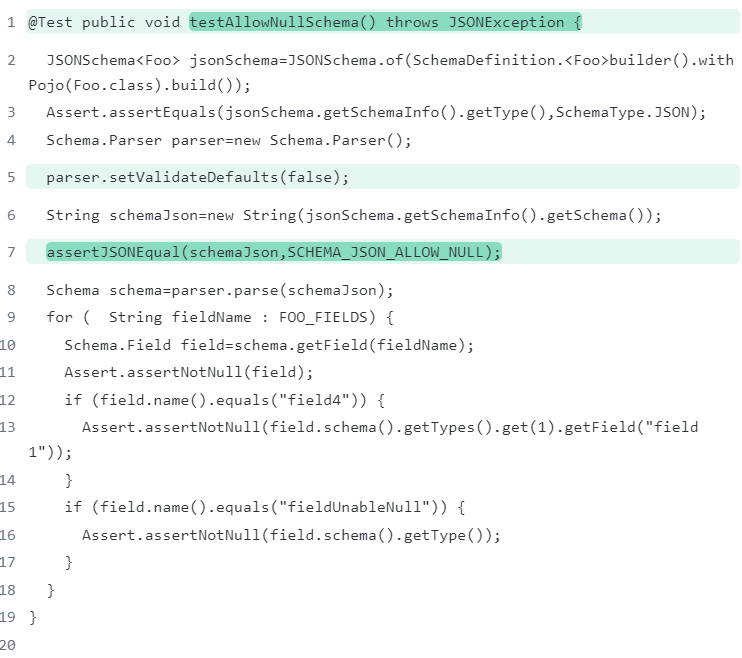} }}%
    \caption{Example of Flaky test (Top) and its developer repair (Bottom) where flakiness is removed by changing assertion and adding exception handling. Our prediction model correctly classifies this flaky test under the Handle Exception and Change Assertion fix category.}%
    \label{fig:HandleExceptionTP}%
\end{figure}

\begin{figure}%
    \centering
    \subfloat[\centering Flaky test]{{\includegraphics[width=9cm]{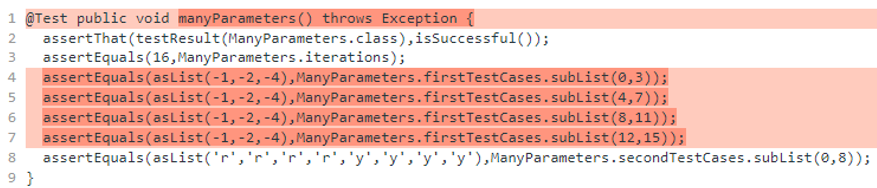} }}%
    \qquad
    \subfloat[\centering Fixed flaky test]{{\includegraphics[width=9cm]{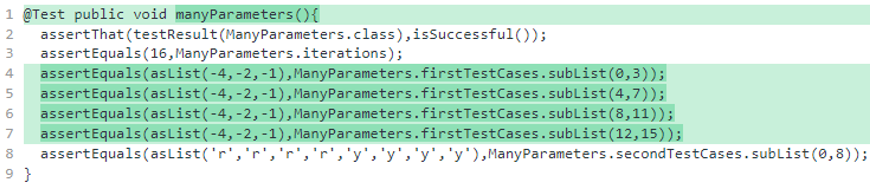} }}%
    \caption{Example of Flaky test (Top) and its developer repair (Bottom) where our prediction model is unable to correctly classify this flaky test under the Handle Exception fix category.}%
    \label{fig:HandleException_example2}%
\end{figure}

Figure~\ref{fig:changedatastructure1} shows a representative example test whose fix category is correctly predicted as \textit{Change Data Structure} (TP). On line 2 of this test code, \textbf{HashMap} is used, which can cause flakiness, as discussed in Section \ref{Approach}. Our model correctly predicts the category and suggests replacing \textbf{HashMap} with \textbf{LinkedHashMap}. 
Similarly, Figure~\ref{fig:HandleExceptionTP} (a) shows another  Flaky test whose fix category is correctly predicted as Handle Exception and Change Assertion (TP). Figure~\ref{fig:HandleExceptionTP} (b) shows the developer fix of the same flaky test where a correct exception handling was added and  ~\textit{assertEquals} was changed to ~\textit{assertJSONEqual} for correctly handling JSON objects. 
The test code shown in Figure~\ref{fig:HandleException_example2} (a) is an example where our prediction model fails to correctly predict the fix category (FN) as ~\textit{Handle Exception}. This test case differs from the typical patterns observed in that fix category because the keyword exception was already present in the method signature. Typically, exception handling is not part of flaky code in that category and an exception or try-catch is added in the code to resolve flakiness. However, in this case, it is the opposite and the pattern does not match other examples in the training set. When we examine the corrected test code shown in Figure~\ref{fig:HandleException_example2} (b), we notice the exception had been removed as part of the fix. However, our automated labeling tool classified this test case as \textit{handle exception} due to the presence of the 'exception' keyword.
% In contrast, the test code in Figure~\ref{fig:changedatastructure2} (top) is an example where the model is unable to predict the correct fix category (FN). As discussed earlier, this test code does not follow the pattern of flaky statements for this fix category and does not contain any specific keywords, such as Hash, Map or Set. To further understand the issue, we examined the developer's fix of this test, as shown in Figure~\ref{fig:changedatastructure2} (bottom). We found that the developer initialized new data structures \textbf{HashSet} on code lines 4 and 6, which resulted in labeling this test with the \textit{Change Data Structure} category by our automated labeling tool. However, since there is no specific data structure in the flaky code that needs to be replaced, the model could not predict the correct fix category, by analyzing only the flaky test code and not the fix.

\begin{figure*}[hbt!]
  \centering
    \includegraphics[width=0.7\linewidth]{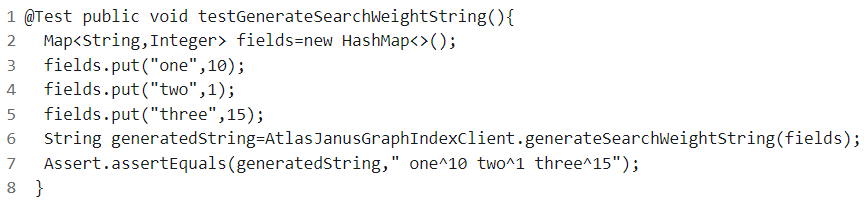}
    \caption{Example of a flaky test where the model incorrectly classifies it in the Change Assertion fix category.}
    \label{fig:changeAssertion_TruePositive}
\end{figure*}

In general, we do not have many examples of FP cases (13 tests on mean for all of the four approaches we have used), which is not surprising as it is one of the strengths of our proposed solution. One sample FP test code is shown in Figure~\ref{fig:changeAssertion_TruePositive}. This test was incorrectly classified as a \textit{Change Assertion} since no change in the assertion type is needed. A potential reason for the misclassification can be related to the long string passed to the assert statement in the code. The pattern of using the long string is similar to cases where flakiness is addressed by changing an assertion to a more customized assertion with shorter inputs, for example using ~\emph{AssertJsonStringEquals} to handle Json Strings. However, here the incorrect use of assert was not the cause for flakiness and thus was not changed.  Although the recommendation was a FP, suggesting a Change in assertion might not be a bad idea after all, not as a fix but as refactoring to create more maintainable tests.

To summarize, we can conclude that the models are correctly learning the patterns that exist in the test code and, in most cases, FP or FN are simply due to lack of data, where without the actual fix or other sources of information about the bug, correct prediction is not possible.
\subsubsection{GPT-Generated Repair of Flaky Tests}
To examine repaired flaky tests generated by GPT, we analyze three prompting methods: (a) a simple prompt without the fix category label, (b) the same prompt but with the label, and (c) adding extra context (sample fixes) to the prompt, i.e., in-context learning (described in Section~\ref{prompt_design}).

\begin{figure*}[hbt!]
  \centering
    \includegraphics[width=1.0\linewidth]{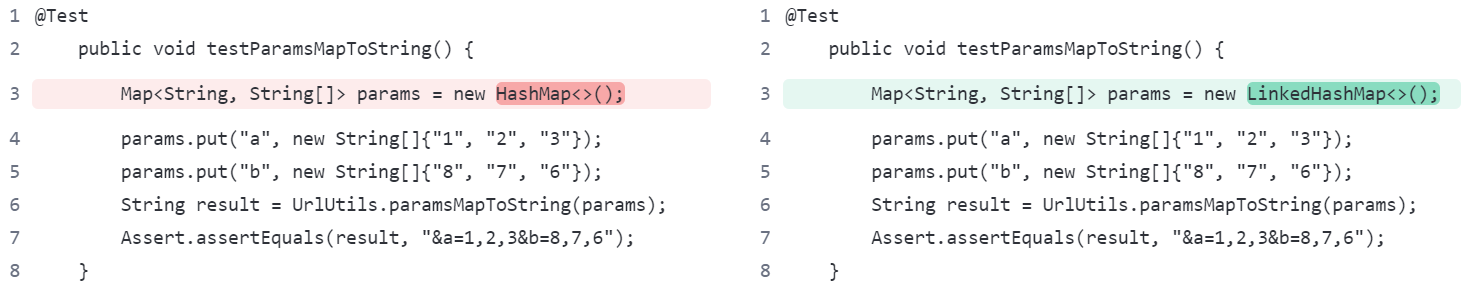}
    \caption{Example of a flaky test (left) and its original developer repair (right). Here flakiness is removed by changing the data structure i.e. replacing HashMap with LinkedHashMap.}
    \label{fig:Data_structure_generation}
\end{figure*}

\begin{figure*}[hbt!]
  \centering
    \includegraphics[width=1.0\linewidth]{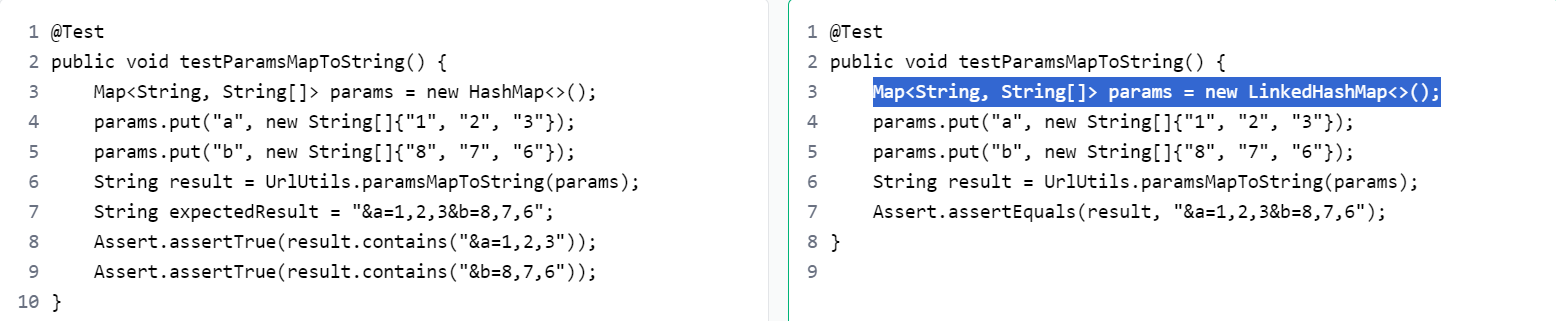}
    \caption{Comparison between GPT-generated repaired flaky tests without a fix category label prompting (left) and with a fix category label provided(right), illustrating GPT's accurate generation of the repaired flaky test when given the fix category label in the prompt. }
    \label{fig:Data_structure_generation2}
    \vspace{-3pt}
\end{figure*}

Figure~\ref{fig:Data_structure_generation} depicts a flaky test and its fix by the original developer. Both UniXcoder and CodeBERT classified this as a \textit{Change Data Structure} fix, replacing HashMap with LinkedHashMap to maintain the elements' order. Figure~\ref{fig:Data_structure_generation2}  shows the results based on prompting methods (a) without the \textit{Change Data Structure} label and (b) with that label. 

When the model was provided with the fix category label, the output matched the developer's repair (Figure~\ref{fig:Data_structure_generation} -- right), achieving a 100\% CodeBLEU Score. However, without the label indicating which statements to fix, the model suggested fixes on the wrong statements unrelated to the issue.

\begin{figure*}[hbt!]
  \centering
    \includegraphics[width=1.05\linewidth]{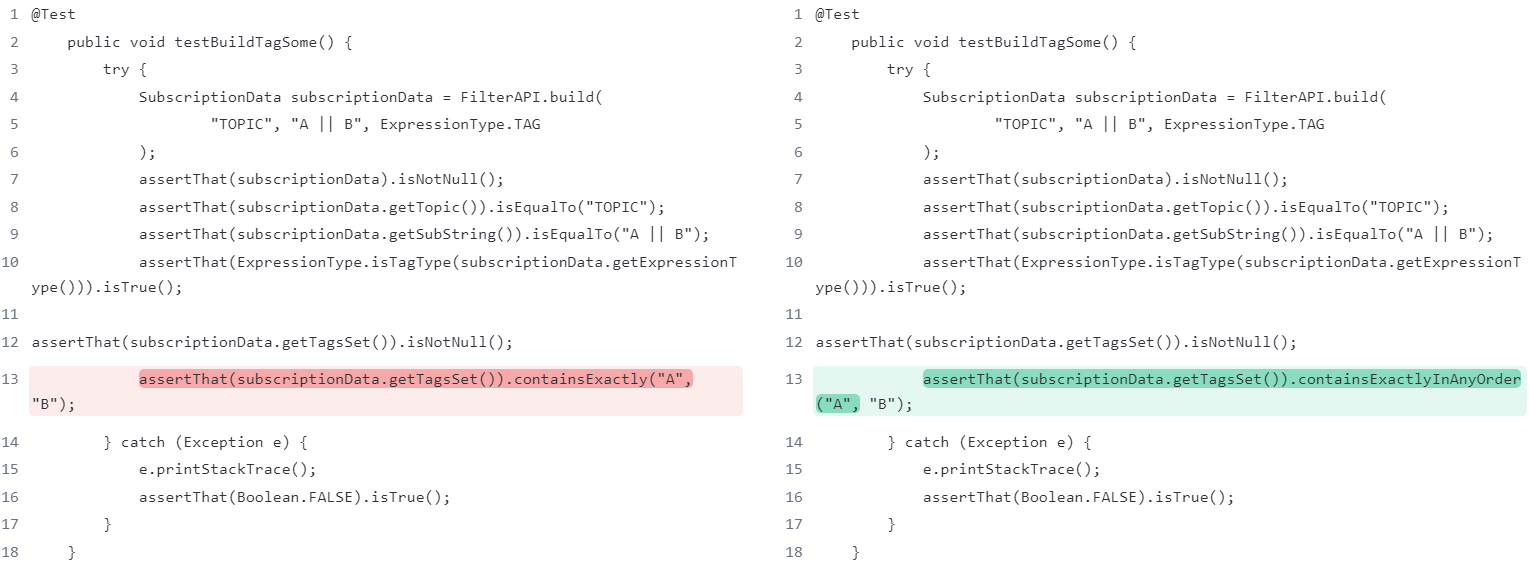}
    \caption{Example of a flaky test (left) and its original developer repair (right). Here flakiness is removed by changing the condition, i.e., replacing .containsExactly() with .containsExactlyInAnyOrder().}
    \label{fig:change_condition_1}
    \vspace{-3pt}
\end{figure*}

\begin{figure}[hbt!]
    \centering
    \subfloat[\centering Repaired flaky test with fix category label.]{{\includegraphics[width=9cm]{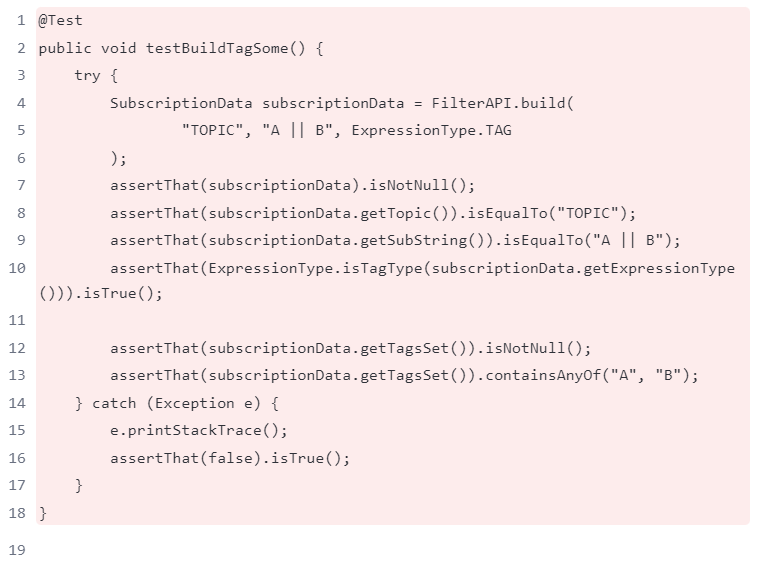} }}%
    \qquad
    \subfloat[\centering Repaired flaky test without fix category label.]{{\includegraphics[width=9cm]{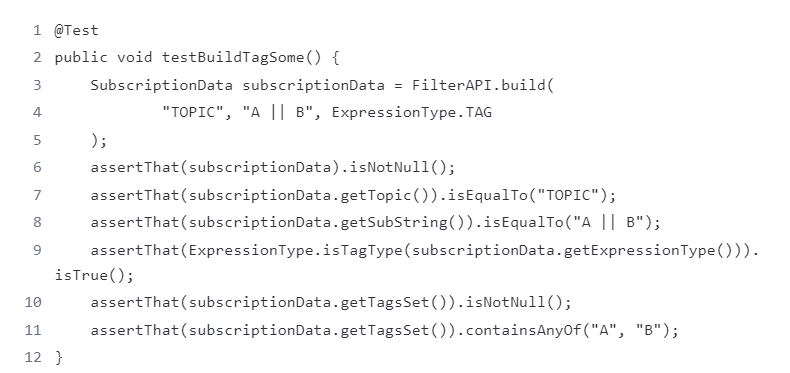} }}%
    \qquad
    \subfloat[\centering Repaired flaky test with in-context learning.]{{\includegraphics[width=9cm]{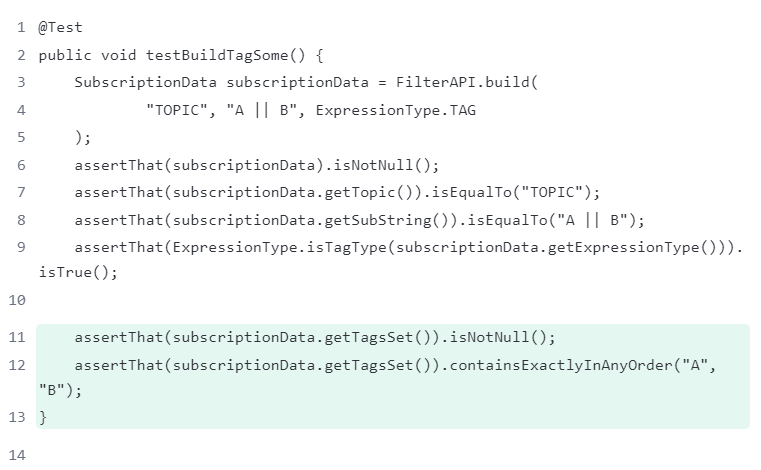} }}%
    \caption{Effect of different prompts on GPT's repaired flaky test generation:
with fix label (Top), without fix label (middle), and with in-context learning (bottom). }
    \label{fig:change_condition_2}%
\end{figure}

Now let's look at another example from the \textit{Change Condition} category, showcased in Figure~\ref{fig:change_condition_1}. Here, the repair modified \textit{.containsExactly()} to \textit{.containsExactlyInAnyOrder()} to fix flakiness. Figure~\ref{fig:change_condition_2} shows the repaired tests with and without the fix category label. It also exemplifies a repair where we provided guidance within the prompt for fixing flaky tests caused by change conditions.

In all generated instances, the generated fix is not an exact match to the original developer's repair. In the first two cases, \textit{.containsExactly()} was replaced with \textit{.containsAnyof()}, which is similar to the ground truth, but does not match exactly. The label-free model wrongly removed the exception handling. However, with the label, the model preserved the structure and attempted to fix the problematic statements.

In the third case, in-context learning, the model correctly replaced \textit{.containsExactly()} with \textit{.containsExactlyInAnyOrder()} based on what model learned from the examples we passed in the prompt. However, it also incorrectly removed proper exception handling, similar to the first instance. Based on our observation, it seems that GPT's accuracy in generating fixes depends on whether the prompt contains the fix category and the number of examples. Future work could improve this accuracy by augmenting the training data and enhancing prompt quality.

%Our analysis indicates that fix category labels enhances automated flaky test repairs, providing clear guidance for developers. Future work could explore improving GPT's accuracy through additional training and better prompt formulation. 
We have included all the generated results in our replication package. 
\subsection{Application of FlakyFix in Practice?}
%How might this approach be used practically? 
In practice, our framework can help fix tests both automatically or with some human intervention. When a test is flagged as problematic (either by a person or other tools), there are two options to fix flakiness: a manual fix by the tester, or running an automated program repair tool.
In either case, our approach is helpful. If the repair is manual, the predicted categories are given to the testers as suggestions so that they can more quickly fix the bug in the test code.
Alternatively, if a tool such as GPT is used to fix the test, the category labels are helpful for prompting, as shown in the previous section.
However, as discussed in Section~\ref{RQ3_Results}, not all GPT-repaired tests are completely correct and will pass upon execution. This is primarily reflected in the CodeBLEU score and the execution results obtained after running a subset of generated tests. We estimate, however, that a large majority of GPT-repaired test cases should pass, their percentage depending on the specific prompt being used. 
A failed generated test will require some manual repair for it to pass. Our analysis using edit distance reveals that, on average, 16\% of tokens need to be changed in a failing test for it to exactly match the passing, original-developer repaired test. Thus, completely fixing flaky test cases does entail some degree of human effort for a minority of repaired test cases and is not entirely free. However, with newer GPT versions and better prompting, we can hope for increasingly reliable fixes and passing test cases in the future.



\section{Threats to Validity}\label{Threats to Validity}
%per-project analysis
%different implementation libraries: Pytorch vs TensorFlow
\subsection{Construct Threats}
%Construct threats to validity are concerned with the extent to which our analyses measure what we claim to be analyzing. 
In our study, the main potential construct validity threat is the possibility of defining incorrect or incomplete labels in RQ1, which can undermine results in RQ2 and RQ3. To reduce this potential threat, we first made the process of labeling systematic and automated with heuristics that are publicly available in the replication package for further evaluation and improvement. In addition, we also manually checked random samples of the labeled tests, to confirm the correctness of the process.~Furthermore, the accuracy of the results that we obtained for our predicted fix category labels are very high for most of the fix categories, as shown in Table~\ref{tab:Results_RQ2}. Given this, we ensured that correct predicted labels are used in different prompts to LLM in RQ3.

Additionally, For RQ2, We did not conduct a per-project analysis in our study due to the limited number of positive instances for all categories of fixes and the large number of projects in our dataset (96). This implies that for each flaky test in the test set, very few flaky tests from the same project are used in the training set, especially for a specific category. We reported overall results across all projects.

Regarding the evaluation metrics in RQ3, we used the CodeBLEU score, a common metric in text generation tasks. However, we did not execute these automated generated tests to verify their functionality. Due to dependencies in open-source projects, executing all these tests is time and computation-intensive.
As a result, we have not confirmed whether the tests with higher CodeBLEU scores will pass or not, or whether they are even executable. However, improvements in CodeBLEU are unlikely to yield worse test cases, in terms of functionality or execution. 
Nevertheless, the inadequacy of the current metrics is a known challenge~\cite{evtikhiev2023out} in this field and future research is needed to reduce such threats.

%we attempted to automate the labeling process of flaky tests with various fix categories by creating a set of heuristics relying on keywords from the \textit{diff} of flaky tests and their fixed version. However, these heuristics may not be perfect and lead to incorrect labeling, for example, if we miss relevant keywords. To ensure that the tool labels the tests in the same way a human would analyze them manually, we manually analyzed the tests before and after the automated labeling process. We have made our labeling tool publicly available in the replication package~\cite{X}.
%While our prediction model offers developers guidance for addressing flaky tests, it is not an all-encompassing solution. A single flaky test might need more types of fixes than our defined categories cover.

%Likewise, the complete fix generated by our GPT model is not currently 100\% accurate. It might mistakenly modify statements that are not causing flakiness. 
\subsection{Internal Threats}
Internal threats to validity are about to make sure the cause-effect relation identified in the study is really there and there is no other explanation (confounding factors). In our study, the claim is that ~\emph{our predicted fix categories will help fix flakiness}. The main confounding factor is about the way we prompt the GPT-3.5 model, when asking to fix a flaky test. Better results might be obtained in the future with additional guidance and more examples in the prompt. Similarly, using GPT-4 would probably lead to even better results but, as discussed in Section~\ref{Repairing_Flaky_Tests_Using GPT_and_Proposed_Fix_Categories}, this would raise, on our data set, possible data leakage issues. 
Additionally, it would be insightful to analyze how often GPT generates a correct fix when provided with an incorrectly predicted fix category. However, as explained in Section~\ref{ref:Dataselection_execution}, we were limited to executing only 35 tests. Within our sample of 35 executable test cases, we encountered only one instance where the fix category was incorrectly predicted, albeit partially correct as one of the two categories was accurately predicted. Therefore, conducting this analysis reliably is not possible with our current data.

%Creating the prompt for different tasks is still quite subjective, which may result in sub-optimal results. However, in our case, a more optimized prompt can only improve the results in favor of our method. 

Another internal validity threat while predicting the labels is the models context size (510 tokens for CodeBERT and 1022 for UniXcoder), which may truncate the source code of some test cases and result in information loss. In our dataset, 49 tests (11\% of the total 562 flaky tests) exceed the token size limit of 510 tokens, and 17 tests  (4\%) exceed 1022 tokens. This could potentially impact the accuracy of our prediction results, especially for those fix categories where we have limited positive examples. Further research should investigate the extent to which this limitation may affects the outcomes of our study.

%Since many projects only have one or two flaky tests, we cannot analyze all fix categories for each project individually. However, if we have more examples from each project for each fix category in the future, we may be able to conduct a per-project analysis.

\subsection{Conclusion Threats}
To reduce potential conclusion validity threats, in RQ2, we used Fisher's Exact test to make sure our conclusions are statistically significant. We also ran Wilcoxon signed-rank test when comparing different prompt types in RQ3. However, there is still a potential conclusion threat in RQ3 results with respect to the inherent randomness of GPT models. Indeed, using the same model and the same prompt can result in different outputs when using higher temperatures. In general, the field of prompt engineering, which also includes the optimization of outputs with several rounds of prompting, is a new field and there are limited precise and reliable guidelines in the literature. Future work will investigate future prompting strategies as they get reported.

\subsection{External Threats}
%External validity threats pertain to the generalizability of study outcomes. To ensure the tests marked as fixed in the IDoFT dataset are indeed fixed and no longer exhibit flakiness, we solely used the developer-approved fixed tests from the IDoFT GitHub repository. Furthermore, 

Our automated labeling tool and the heuristics are defined for Java tests. We have not evaluated our tool on other programming languages but it can be adapted to other languages and new projects by incorporating adapted keywords and rules. Additionally, CodeBERT,  UniXcoder and GPT models are pre-trained on multiple programming languages.

\section{Related Work}\label{Related Work}
%Flaky Test 
%Fixing Flaky Tests
%Predicting Flaky Tests Fixes and root Causes Categories 

% opening paragraph: excluding: generic program repair. - test smells
In this section, we provide an overview of the most related work to predicting flaky test fix categories. There are two areas of research that we deliberately exclude, since they are less related to our study. 
The first category that we do not include is generic program repair studies~\cite{goues2019automated}. Although, fixing test code can be seen as a similar task to program repair, there are fundamental differences between the two areas which are mainly due to the patterns of fixes. A generic code repair can come in many flavors and patterns and thus predicting a category of fix is a daunting task. In addition, there are far more data available for generic program repair (e.g., literally any code fix in GitHub) than for fixing test flakiness. Therefore, we do not discuss individual papers that study generic program repair.
The second category that we exclude here are studies that analyze test smells and their patterns~\cite{garousi2018smells}. Although these works are more relevant since they are about issues in the test code, given that they are again generic and not related to flakiness, we exclude them.

The first set of relevant papers to our study are those that specifically address categories or taxonomies of fixes or fix patterns. Although they are not about flaky test fixes or not even about test cases, these studies have nevertheless a similar problem to solve: categorizing code fixes or causes and predicting them. 
Ni et al~\cite{thung2012automatic} proposed an approach to classify bug fix categories using a Tree-based Convolutional Neural Network (TBCNN). To do so, they used the \textit{diff} of the source code before and after the fix at the Abstract Syntax Tree (AST) level to construct fix trees as input for TBCNN.
In contrast, our objective is to develop a classifier for fix categories that only use flaky test code to make predictions, without relying on the repaired versions or production code.

Predicting the cause of flakiness is another category that is relevant to our work. Akli et al. proposed a novel approach~\cite{akli2023flakycat} for classifying flaky tests based on their root cause category using CodeBERT and FSL. They manually labeled 343 flaky tests with 10 different categories of flaky test causes suggested by Luo et al~\cite{luo2014empirical} and Eck et al~\cite{eck2019understanding}. However, they reported the prediction results for only four categories of causes, namely Async wait, Concurrency, Time, and Unordered collection, as they lacked sufficient positive examples to train an ML-based classifier for the other categories. They achieved the highest prediction results for the Unordered collection cause, with a precision of 85\% and a recall of 80\%. In a recent survey on flaky tests~\cite{parry2021survey, lam2019root},  
Parry et al. mapped three different causes of flakiness with their potential repairs. For instance, they suggested adding or modifying assertions in the test code to fix the flakiness caused by concurrency. Predicting fix categories provides testers with a specific direction for modifying the code. Knowing that concurrency is causing flakiness in a given test is not as useful to testers as they do not know which part of the code is causing the issue. However, knowing that changing assertions can eliminate flakiness gives them more practical guidance.
Other researchers reported  studies~\cite{zheng2021research,lam2019root} to identify flakiness causes, impacts, and existing strategies to fix flakiness. However, none of them attempted to categorize the tests based on their fixes.
Research on repairing flakiness and categorizing fix categories is still in its early stages, with much of the work focused on fixing order-dependent flaky tests. For example, iPFlakies~\cite{wang2022ipflakies} and iFixFlakies~\cite{shi2019ifixflakies} propose methods for addressing order-dependent flakiness, while Flex~\cite{dutta2021flex} automates the repair of flaky tests caused by algorithmic randomness in ML algorithms. Parry et al.\cite{parry2022developer} proposed a study in which different categories of flaky test fixes were defined and 75 flaky commits were categorized based on these categories. However, the categorization was performed through manual inspection of the commits, code diffs, developer comments, and other linked issues outside the scope of the test code. In another study of flaky tests in the Mozilla database\cite{eck2019understanding} tests were categorized based on their origin (i.e., the root cause of the flakiness, whether in the source or test code) and developers' efforts to fix them.
The need for automated fixing of flaky tests and conducting quick small repairs was strongly suggested in a recent survey on supporting developers and testers in addressing test flakiness~\cite{gruber2022survey}. Our study aims to address these gaps by developing an automated solution to categorize flaky tests based on well-defined  fix categories, relying exclusively on test code, providing practical guidance for both manual and automated repair. We also provide a way to automatically label flaky tests with selected, common fix categories to train context-specific prediction models. 
%Additionally, we propose a prediction model that identifies the fix category that can be applied to eliminate flakiness for a given flaky test. This approach can provide developers with a quick direction for repairing flaky tests.

Although we do not review the generic automated program repair literature, as explained, we do however discuss related work where LLMs are used for automatic program repair, since it is very relevant to RQ3 from a technical standpoint. Recent advancements in LLMs, exemplified by ChatGPT \cite{fu2023chatgpt}, Copilot \cite{nguyen2022empirical}, and similar models tailored for code, have significantly advanced this field. These models exhibit adeptness in code generation and are generating considerable interest among researchers striving to enhance automatic program repair methodologies. 

Ribeiro et al. \cite{ribeiro2023gpt} employed GPT-3 to automatically rectify bugs in 1318 buggy OCaml programs, achieving a repair rate of 39\%. Julian et al. \cite{prenner2022can}, using OpenAI's Codex model, repaired bugs in Java and Python programs, noting a higher repair rate for Python. Their utilization of prompts and hints improved Codex's bug identification, yielding slightly enhanced results in repairing Java programs. In another recent study by Chunqiu et al.~\cite{xia2023automated}, different LLMs like Codex, InCoder, CodeT5 and GPT-2 are investigated for program repair in Java, Python, and C programs. 

Additionally, Mashadi et al. \cite{mashhadi2021applying} and Xia et al. \cite{xia2022less} utilized CodeBERT for automated program repair on datasets like ManySStuBs4J and Defects4J, respectively, achieving state-of-the-art results in bug repair for smaller code segments due to CodeBERT's input limitation of up to 512 tokens.

However, while existing research has focused on rectifying bugs in source code, little attention has been directed towards addressing flakiness within test code itself. Our study marks the first attempt to fully automate flaky test repairs using GPT, complemented by providing fix category information to pinpoint flakiness related statements in the code and enhance the repair capability of the GPT model.
 
\section{Conclusion and Future Work}\label{Conclusion}
In this paper, we proposed a framework to (a) categorize flaky tests according to practical fix categories based exclusively on test code and (b) use the fix categories to generate fully automated repairs of the flaky tests with language models.  Given the limited availability of open-source datasets for flaky test research, our generated labeled data can also be valuable for future research, particularly for ML-based prediction and repair of flaky tests. We evaluated two language models, CodeBERT and UniXcoder, with and without Few-Shot Learning, and found that the UniXcoder model outperformed CodeBERT in predicting most fix categories.  We also showed that the category information, complemented with in-context learning can greatly help a large language model (LLM), namely GPT-3.5 in our case, better fix the flakiness in test code.~Lastly, we executed a sample of GPT-repaired flaky tests and estimated that a large percentage (roughly between 51\% and 83\%) of these tests can be expected to pass. For the generated tests that fail, an average of 16\% of the code needs further modification for them to pass.
In the future, we plan to expand our flaky dataset with more data. We also plan to expand our heuristics for automatically labeling flaky tests to other programming languages. Replicating the study with other LLMs (such as GPT-4.0) will also be made possible with an updated dataset.
Additionally, a significant percentage of flaky tests may stem from problems in the production code, which cannot be addressed by black-box models. Therefore, in the future, we need to develop lightweight and scalable approaches for accessing and fixing the production code when dealing with test flakiness.
\section*{Acknowledgement}
This work was supported by a research grant from Huawei Technologies Canada, Mitacs Canada, as well as the Canada Research Chair and Discovery Grant programs of the Natural Sciences and Engineering Research Council of Canada (NSERC). The experiments conducted in this work were enabled in part by WestGrid (https://www.westgrid.ca) and Compute Canada (https://www.computecanada.ca). Dr. Hemmati's research is partly supported by NSERC and Alberta Innovates.
\bibliographystyle{unsrt}
%\bibliographystyle{IEEEtran}
\bibliography{IEEEabrv,references.bib}
\section*{Authors' Biographies}
\vspace{0pt}
\begingroup
\setlength{\intextsep}{-1.5pt}
\begin{wrapfigure}{l}{36mm} 
    \includegraphics[width=1.5in,height=1.87in,clip,keepaspectratio]{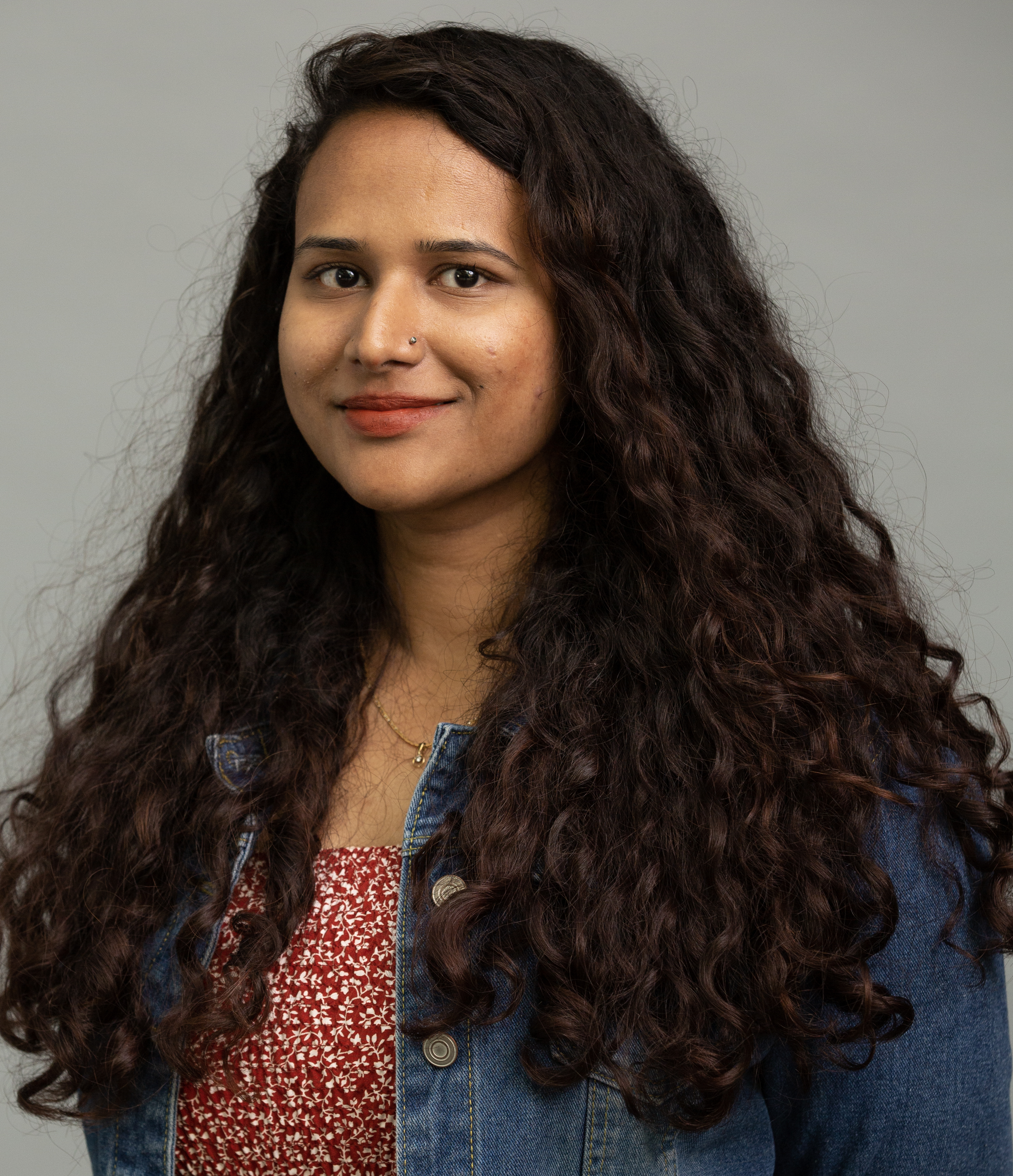}
  \end{wrapfigure}\par
  \textbf{Sakina Fatima} is a final-year PhD candidate at the School of Electrical Engineering and Computer Science at the University of Ottawa and a member of Nanda Lab. Sakina has worked in collaboration with Microsoft Research, IBM and Huawei. She obtained an Erasmus Mundus Joint Masters degree in Dependable Software Systems from the University of St Andrews, United Kingdom and Maynooth University, Ireland. In 2019, she was awarded the French Government Medal and the National University of Ireland prize for distinction in collaborative degrees. Her research interests include automated software testing, generative AI, large language model and applied machine learning.\par
\endgroup

\vspace{8pt}

\begingroup
\setlength{\intextsep}{-1.5pt}
\begin{wrapfigure}{l}{36mm} 
    \includegraphics[width=2.7in,height=1.73in,clip,keepaspectratio]{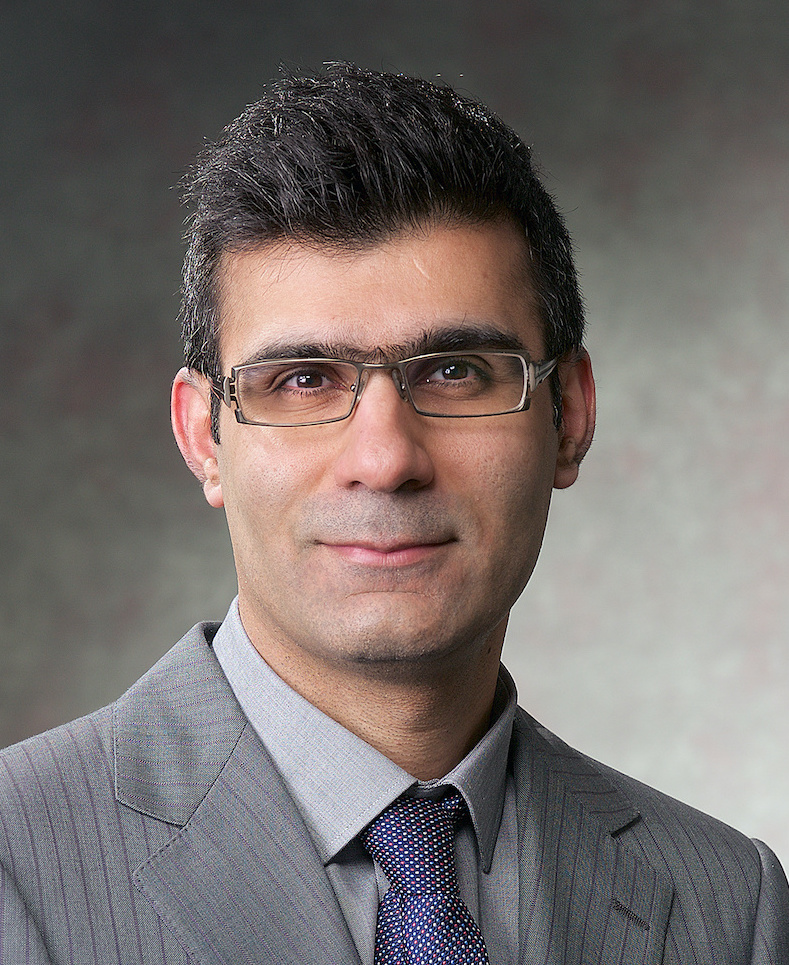}
  \end{wrapfigure}\par
  \noindent\textbf{Dr. Hadi Hemmati}  is an associate professor at the electrical engineering and computer science department, at York University. Previously he was an associate professor at the electrical and software engineering department at the University of Calgary, AB, Canada. In the past, he was also an assistant professor at the University of Manitoba, and a postdoctoral fellow at the University of Waterloo, and Queen’s university. He received his Ph.D. from the University of Oslo, Norway. His main research interests are automated software engineering (with a focus on software testing, debugging, and repair), and trustworthy AI (with a focus on robustness and explainability). His research has a strong focus on pragmatic software/ML solutions for large-scale systems and empirically investigating them in practice. He has been a PI on multiple industry research projects in different domains such as IT, aviation, insurance, urban development, fintech, and beyond.\par
\endgroup

\vspace{8pt}

\begingroup
\setlength{\intextsep}{-1.5pt}
\begin{wrapfigure}{l}{36mm} 
    \includegraphics[width=2in,height=1.5in,clip,keepaspectratio]{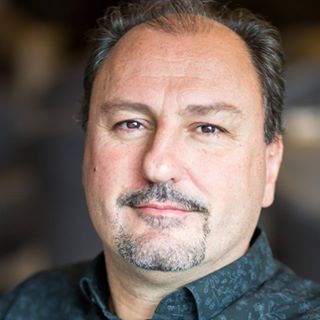}
  \end{wrapfigure}\par
  \noindent\textbf{Dr. Lionel C. Briand}  is professor of software engineering and has shared appointments between (1) The University of Ottawa, Canada, and (2) The Lero SFI Centre---the national Irish centre for software research---hosted by the University of Limerick, Ireland. In collaboration with colleagues, for over 30 years, he has run many collaborative research projects with companies in the automotive, satellite, aerospace, energy, financial, and legal domains. Lionel has held various engineering, academic, and leading positions in seven countries.  He currently holds a Canada Research Chair (Tier 1) on "Intelligent Software Dependability and Compliance" and is the director of Lero, the national Irish centre for software research. Lionel was elevated to the grades of IEEE Fellow and ACM Fellow for his work on software testing and verification. Further, he was granted the IEEE Computer Society Harlan Mills award, the ACM SIGSOFT outstanding research award, and the IEEE Reliability Society engineer-of-the-year award. He also received an ERC Advanced grant in 2016 on modelling and testing cyber-physical systems, the most prestigious individual research award in the European Union and was elected a fellow of the Academy of Science, Royal Society of Canada in 2023. More details can be found at: http://www.lbriand.info.\par
\endgroup
\end{document}